\documentclass{iopart}
\usepackage{graphicx,epsfig}
\usepackage{bm}
\usepackage{iopams}
\newcommand{\be}{\begin{equation}}
\newcommand{\ee}{\end{equation}}
\newcommand{\ba}{\begin{eqnarray}}
\newcommand{\ea}{\end{eqnarray}}
\newcommand{\bse}{\numparts}
\newcommand{\ese}{\endnumparts}

\newcommand{\bbq}{\begin{quote}}
\newcommand{\eeq}{\end{quote}}
\newcommand{\RR}{{}^3{\cal{R}}}

\newcommand{\HH}{{\cal{H}}}
\newcommand{\KK}{{\cal{K}}}
\newcommand{\PP}{{\cal{P}}}

\newcommand{\Ome}{\hat\Omega_{e}}
\newcommand{\rhom}{\rho^{(m)}}
\newcommand{\rhomi}{\rho_0^{(m)}}
\newcommand{\rhoqm}{\rho_q^{(m)}}
\newcommand{\rhoqmi}{\rho_{q0}^{(m)}}

\newcommand{\rhoe}{\rho^{(e)}}
\newcommand{\rhoei}{\rho_0^{(e)}}
\newcommand{\rhoqe}{\rho_q^{(e)}}
\newcommand{\rhoqei}{\rho_{q0}^{(e)}}

\newcommand{\Ommq}{\Omega_q^{m}}
\newcommand{\Ommqi}{\Omega_{q0}^{m}}
\newcommand{\Omeq}{\Omega_q^{e}}
\newcommand{\Omeqi}{\Omega_{q0}^{e}}
\newcommand{\Omkq}{\Omega_q^{k}}
\newcommand{\Omkqi}{\Omega_{q0}^{k}}
\newcommand{\Dm}{\delta^{m}}
\newcommand{\De}{\delta^{e}}
\newcommand{\Dk}{\delta^k}
\newcommand{\Dh}{\delta^{\HH}}
\newcommand{\Dj}{\delta^{J}}
\newcommand{\Omm}{\hat\Omega_{m}}

\newcommand{\dd}{{\rm{d}}}

\begin{document}
\title{Interactive mixture of inhomogeneous dark fluids driven by dark energy: a dynamical systems analysis.}
\author{Germ\'an Izquierdo${}^\dagger$, Roberto C. Blanquet-Jaramillo${}^\dagger$ and Roberto A. Sussman${}^\ddagger$}
\address{${}^\ddagger$ Facultad de Ciencias, Universidad Aut\'onoma del Estado de M\'exico, Toluca 5000, Instituto literario 100, Edo. Mex.,M\'exico.\\
${}^\ddagger$ Instituto de Ciencias Nucleares, Universidad Nacional Aut\'onoma de M\'exico (ICN-UNAM),A. P. 70--543, 04510 M\'exico D. F., M\'exico.
 }
\ead{gizquierdos@uaemex.mx}
\date{\today}
\begin{abstract}
We examine the evolution of an inhomogeneous mixture of non-relativistic pressureless cold dark matter (CDM), coupled to dark energy (DE) characterised by the equation of state parameter  $w<-1/3$, with the interaction term proportional to the DE density.  This coupled mixture is the source of a spherically symmetric Lema\^\i tre--Tolman--Bondi (LTB) metric admitting an asymptotic Friedman--Lema\^\i tre--Robertson--Walker (FLRW) background. Einstein's equations reduce to a 5-dimensional autonomous dynamical system involving quasi--local variables related to suitable averages of covariant scalars and their fluctuations. The phase space evolution around the critical points (past/future attractors and five saddles) is examined in detail. For all parameter values and both directions of energy flow (CDM to DE and DE to CDM) the phase space trajectories  are compatible with a physically plausible early cosmic times behaviour near the past attractor. This result compares favourably with mixtures with the interaction driven by the CDM density in which conditions for a physically plausible past evolution are more restrictive.  Numerical examples are provided describing the evolution of an initial profile that can be associated with idealised structure formation scenarios.
\end{abstract}
\pacs{98.80.-k, 04.20.-q, 95.36.+x, 95.35.+d}

\section{Introduction}
Current observational evidence supports the existence of an accelerated cosmic expansion, likely driven by an unknown form of matter--energy, generically denoted ``dark energy'' (DE), and usually described by suitable scalar fields or (phenomenologically) as a fluid with negative pressure \cite{copeland, wmap, plank}. Observations also point to the existence of cold dark matter (CDM) clustering around galactic halos, usually described in cosmological scales by pressure--less dust, while ordinary visible matter (baryons, electrons and neutrinos) and photons (radiation) comprise less than 5\% of the total contents of cosmic mass--energy.

While both dark sources only interact with ordinary matter and radiation through gravitation, it is very reasonable to assume that there is some form of interaction between them. This assumption cannot be ruled out, given our ignorance on the fundamental nature of these sources. In fact, potentially useful information on the primordial physics behind dark sources may emerge by fitting various assumptions of such interactions to observational data, given the fact that interactive DE and CDM is consistent with  the dynamics of galaxy clusters \cite{abdalla} and the integrated Sachs-Wolfe effect \cite{ol3}. Several models of coupled dark sources can also be found in the literature motivated by particle physics, thermodynamics, etc. \cite{copeland,boe08}.

Observations also suggest that at sufficiently large scales the Universe is well described by linear perturbations of all sources (dark and visible) in an homogeneous Friedman--Lema\^\i tre--Robertson--Walker (FLRW) background metric, with non--linear dynamics (whether Newtonian or relativistic) needed to explain the observed local structure \cite{linperts}. The interplay of local and cosmic dynamics at all scales must comply with the observed anisotropy of the Cosmic Microwave Background (CMB) \cite{copeland, wmap, plank}. Evidently, this dynamics depends on the assumptions made on DE and CDM, which leads to a model dependent power spectrum that should be contrasted with observations at large scales and in structure formation (from data and from numerical simulations). The observed data should provide interesting constraints on assumptions about the dark sources. In particular, different DE models have been considered in the linear order perturbation scheme in the literature \cite{ol, ol2, Maar, Gavela}.

Conventionally, structure formation scenarios are studied by non--linear Newtonian dynamics (analytically \cite{newt1,newt2} and through numerical simulations \cite{newt3}, see review \cite{newt4}), since CDM is assumed to be practically pressure--less and DE can be modelled (or approximated) by a cosmological constant. However, once we assume a fully dynamical DE source with non--trivial pressure and non--trivial interaction with CDM, it is necessary to utilise General Relativity (whether perturbatively or not) to obtain a valid description of its evolution, since Newtonian gravity can (at best) mimic sources with pressure when adiabatic conditions are assumed (see discussion in \cite{nonewt}). While considering a scalar field is the most common approach to dynamical DE, a phenomenological description by means of fluids with negative pressure can also be useful. Ideally, fully relativistic inhomogeneous DE and CDM interacting sources should be examined through high power numerical relativistic codes (whether assuming a continuous modelling or N--body simulations). However, since the latter codes are in their early development stages \cite{Relsim1,Relsim2,Relsim3,Relsim4}, we can resort to more idealised (yet still relativistic and non--perturbative) description by means of inhomogeneous exact solutions of Einstein's equations.

In most of the literature the term ``Lema\^\i tre--Tolman--Bondi (LTB) models'' is broadly understood to denote spherically symmetric exact solutions endowed with the LTB metric and associated with a pure dust source \cite{LTB}. These solutions have been extensively studied (with zero and nonzero cosmological constant) and used in a wide range of astrophysical and cosmological modelling (see extensive reviews in \cite{ltbrev1,ltbrev2,ltbrev3,ltbrev4}). In particular, a better understanding of their theoretical properties follows by describing their dynamics in terms of ``quasi--local scalars'' \cite{RadAs,RadProfs,suss13a,suss13b} (to be denoted henceforth as ``q--scalars''), which are related to averages of standard covariant scalars and satisfy FLRW dynamical equations and scaling laws \cite{suss13a}.

Since the deviation from a homogeneous FLRW background can be uniquely determined (in a covariant manner \cite{suss13a}) by fluctuations that relate the q--scalars and the standard covariant scalars, the full dynamics of Einstein's equations is equivalent to the dynamics of the q--scalars and their fluctuations (see discussion in \cite{suss13b}). The q--scalars and their fluctuations allow for a consistent dynamical systems study of the models (with zero cosmological constant in \cite{suss08,sussmodes} and nonzero in \cite{izsuss10}). An important theoretical connection with cosmological perturbation theory follows from the fact that the fluctuations of the q--scalars provide an exact analytic (and covariant) generalisation of gauge invariant cosmological perturbations in the isochronous gauge \cite{sussmodes,suss15}.

It is less known that LTB metrics admit energy momentum tensors with nonzero pressure in a comoving frame. For a perfect fluid the pressure must be uniform (zero pressure gradients), which allows to interpret the source as a mixture composed by a homogeneous DE fluid interacting with inhomogeneous dust representing CDM (see \cite{suss05}). For fluids with anisotropic pressure, the latter supports non--trivial pressure gradients, leading to a similar description in terms of q--scalars and their fluctuations as in pure dust LTB models. For anisotropic fluids the anisotropy of the pressure can be related to the fluctuation of the q--scalar associated with the isotropic pressure. Since setting up fluid mixtures is possible and a wide variety of equations of state are admissible, LTB metrics with these sources have been used to model inhomogeneous mixtures of DE and CDM \cite{sussQL,suss09}.

In order to extend earlier work in \cite{sussQL,suss09} and to explore a generalisation of previous work in \cite{izsuss10} that considered DE as a cosmological constant, we studied recently  \cite{izsuss17} a dynamical systems analysis of an LTB interactive mixture of CDM (dust) and DE (fluid with $p/\rho=w,\,\,w=\hbox{const.}<-1/3)$, under the assumption of an interaction driven by CDM: {\it i.e.} the interaction term $J$ is proportional (via a dimensionless constant $\alpha$) to the CDM dust density.

In the present paper we undertake a dynamical systems analysis of a similar (yet qualitatively different) configuration: we assume the same EOS for CDM and DE, but with the interaction now driven by DE, with $J$ now proportional to the DE density. As we show along the paper, the resulting evolution is qualitatively different in both cases.
While the phase space of both (CDM or DE driven) mixtures contains as critical points five saddles and past/future attractors. The attractors have very different properties:
\begin{itemize}
\item The CDM driven mixture examined in \cite{izsuss17}:  the phase space position of the past attractor (Big Bang) depended on the parameters $\alpha$ and $w$. For $\alpha>0$ (energy transfer from DE to CDM) the past attractor describes a well behaved CDM dominated scenario. However, for $\alpha<0$ (energy transfer from CDM to DE) the DE density became negative in phase space regions around  the past attractor (irrespective of initial conditions), thus signalling an unphysical past evolution that is inconsistent with all observational data. This problem was already noticed in the literature with this type of CDM-DE mixtures based on FLRW metrics \cite{copeland,boe08}.
\item The DE driven mixture examined here. As opposed to the system in \cite{izsuss17}, the past attractor is now fixed and the future attractor depends on the choice of parameters $\alpha$ and $w$. Thus, regardless of the sign of $\alpha$ (directionality of interaction energy flow), the past evolution is now a physically plausible CDM dominated scenario (compatible with observations) while the position of the future attractor describes various plausible DE dominated scenarios that depend on the parameter choices. For $\alpha<0$ the future attractor is unphysical (CDM density becomes negative). In this case, the phase space evolution must be appropriately restricted.
\end{itemize}
Since all observations examine cosmic evolution along our past null cone, the physical plausibility of the models must be determined primarily from their past phase space evolution (the future evolution is much more open to speculation). Hence, a comparison between our results and those of \cite{izsuss17} suggests that the DE driven mixture should be favoured, as it exhibits a wider parameter consistency with a past evolution compatible with observations.

The plan of the article is summarised as follows. In section \ref{genLTB} we present the q--scalar formalism to set up the evolution equations of for an LTB metric whose source is a mixture of CDM and DE with the coupling term proportional to the DE density. The resulting dynamical system and the characteristic features of its phase space are introduced in section \ref{dynsys}. In section \ref{criticalpointsLTB} we classify the critical points for different ranges of the free parameters. In section \ref{numerical} we solve numerically the dynamical system for initial conditions corresponding to three different choices of the free parameters, leading to three different types of evolution. The associated phase space trajectories  and radial profile evolution are displayed in detail. Finally, in section \ref{conclusions} we summarise our findings findings and compare our results with those of \cite{izsuss17}.

\section{LTB spacetimes, q--scalar variables and coupled dark energy model}\label{genLTB}

Following the methodology described in \cite{sussQL,suss09} (summarised in \cite{izsuss17}), we consider an LTB metric in a comoving frame given by (we choose units with $c=1$)
\begin{equation} ds^2 = -dt^2 +\frac{R'^2\,dr^2}{1-K}+R^2[d\theta^2+\sin^2\theta\,d\phi^2],\label{LTB}\end{equation}
where $R=R(t,r)$, \, $R'=\partial R/\partial r$,\, $K=K(r)$, with the total energy--momentum tensor given by
\be T^{ab} = \rho u^a u^b + p h^{ab} + \Pi^{ab},\quad u^a=\delta^a_t,\quad h^{ab}=g^{ab}+u^au^b,\label{Tab}\ee
where $\rho(t,r)$ and $P(t,r)$ are the total energy density and isotropic pressure, while the traceless anisotropic pressure tensor is $\Pi^a_b = \PP(t,r)\times \hbox{diag}[0,-2,1,1]$. Since we are interested in describing  a CDM and DE fluid mixture, we assume the following decomposition of (\ref{Tab})
\be \fl T^{ab}=T_{(m)}^{ab}+T_{(e)}^{ab}\quad\Rightarrow\quad \rho=\rho^{(m)}+\rho^{(e)},\quad p= p^{(m)}+p^{(e)},\quad \PP= \PP^{(m)}+\PP^{(e)},\label{mixture}\ee
where the indices ``$(m)$'' and ``$(e)$'' (whether above or below) respectively denote the CDM and DE mixture components (we adopt this convention henceforth).
The total energy--momentum tensor is conserved:  $\nabla_b T^{ab}=0$, but the decomposition above leads to the  conservation law for the mixture components
\be
\nabla_b T_{(m)}^{ab}= j^a=-\nabla_b T_{(e)}^{ab},
\ee
where $j^a$ is the coupling current that characterises the interaction of both sources.  In order to keep the symmetry of the metric, we assume that this current is a vector parallel to the 4--velocity, so that $j_a=Ju_a$ and $h_{ca}j^a=0$ hold. The projection along $u^a$ is
\begin{equation}u_a \nabla_b T_{m}^{ab}= J =- u_a \nabla_b T_{e}^{ab}.
\label{uTab_cons_J}\end{equation}
while the spatially projected conservation equation $h_{ac}\nabla_bT^{ab}=0$ holds for all the evolution.

We have now five state variables $A=\rho^{(m)},\,\rho^{(e)},\,p^{(m)},\,p^{(e)},\,J$ which depend on $(t,r)$. As shown in \cite{sussQL,suss09,izsuss17}, we associate to each of them a q--scalar $A_q$ and a fluctuation $\delta^A$ by the following rule
\footnote{Notice that $A_q$ is related to the proper volume average of $A$ with weight factor $\sqrt{1+K}$. See comprehensive discussion in \cite{suss13a}}
\be A \mapsto A_q =\frac{\int_0^r{A\,R^2\,R'\,dx}}{\int_0^r{\,R^2\,R'\,dx}},\qquad \delta^A=\frac{A-A_q}{A_q}=\frac{A'_q/A_q}{3R'/R},\label{qmaps}\ee
where the lower bound of the integrals above $x=0$ marks a symmetry centre such that $R(t,0)=\dot R(t,0)=0$, with $\dot R=u^a\nabla_a R=\partial R/\partial t$. In particular, it is straightforward to show (see \cite{suss09}) that
\be \fl p_q^{(m)} =p^{(m)}-2\PP^{(m)},\quad \delta_{(m)}^p=2\PP^{(m)},\qquad p_q^{(e)} =p^{(e)}-2\PP^{(e)},\quad \delta_{(e)}^p=2\PP^{(e)}.\label{pps}\ee
Other covariant scalars associated with (\ref{LTB}) and (\ref{Tab}) are the Hubble expansion scalar $\HH=(1/3)\nabla_a u^a=(R^2R')\,\dot{}/(R^2R')$ and the spatial curvature $\KK=(1/6)\RR=2(KR)'/(R^2R')$, where $\RR$ is the Ricci scalar of constant $t$ hypersurfaces. Their respective q--scalars are given by
\be \HH_q = \frac{\dot R}{R},\qquad \KK_q = \frac{K}{R^2}.\label{HKq}\ee
while for the interaction term we have $J=J_q(1+\Dj)$, whose dependence on state variables will be determined further ahead.

For the mixture (\ref{mixture}) to describe CDM and DE we will choose the following equations of state (EOS) (the same as in \cite{izsuss17}):
\ba \hbox{CDM (dust):}\qquad\qquad\quad p^{(m)}=0\quad\Rightarrow\quad \delta_{(m)}^p=0,\label{eoscdm}\\
\hbox{DE (barotropic fluid):}\qquad p^{(e)}=w\rho^{(e)}\quad\Rightarrow\quad \delta_{(e)}^p=\delta_{(e)}^\rho,\label{eosde}\ea
where we have assumed that $w=p^{(e)}/\rho^{(e)}<-1/3$ is a constant. Given the EOS's (\ref{eoscdm}) and (\ref{eosde}), we will adopt the following convention
\be \delta^m\equiv \delta_{(m)}^\rho,\qquad \delta^e\equiv \delta_{(e)}^\rho.\ee
Notice that (\ref{pps}) and (\ref{eosde}) imply that only the DE source contributes to the anisotropic pressure: $\PP=\PP^{(e)}=\delta_{(e)}^p/2$.

As shown in \cite{sussQL,suss09,izsuss17} Einstein's equations reduce to the following system of evolution equations
\footnote{The system (\ref{evHH_q2})--(\ref{evDh_q2}) can be used to determine the metric coefficients $R$ and $R'$ in (\ref{LTB}) by supplying the following two evolution equations $\dot R = R\,\HH_q,\,\,\, \dot \Gamma = \Gamma \HH_q\Dh$ with $\Gamma=R'/R$, which follow from the first equation in (\ref{HKq}) and the second equation in (\ref{qmaps}) for $A=\HH$.}
\bse\label{eveqs_q2}\ba
\dot\HH_q &=& -\HH_q^2 -\frac{\kappa}{6}\,\left[\rhoqm+(1+3\,w\,)\rhoqe\right]\,,\label{evHH_q2}\\
\dot\rhoqm &=& -3{\HH}_q\,\rhoqm+J_q,\label{evm_q2}\\
\dot\rhoqe &=& -3{\HH}_q\left( 1+w\right)\rhoqe-J_q,\label{eve_q2}\\
\dot\delta^m &=& -3\HH_q\,\left(1+\Dm\right)\Dh-\frac{J_q}{\rhoqm}\left(\Dm-\Dj\right),\label{evDm_q2}\\
\dot\delta^e &=& 3\HH_q\,[w\,\De-\left(1+w+\De\right)\Dh] -\frac{J_q}{\rhoqe}\left(\De-\Dj\right),\label{evDe_q2}\\
\dot\Dh &=& -\HH_q\Dh\left(1+3\Dh\right)+\nonumber\\
&&\frac{\kappa}{6\HH_q}
\left[\rhoqm\,\left(\Dh-\Dm\right)+(1+3w)\rhoqe\,\left(\Dh-\De\right)\right],\label{evDh_q2}
\ea\ese
together with the algebraic constraints
\ba
\HH_q^2=\frac{\kappa}{3}\left(\rhoqm+\rhoqe\right)-\KK_q,\label{cHam2}\\
 2\HH_q^2\Dh =\frac{\kappa}{3}\left(\rhoqm\delta^m+\rhoqe\delta^e\right)-\KK_q\Dk.\label{cHam2b}
\ea
where $\kappa=8\pi/3$, (\ref{cHam2b}) follows from (\ref{cHam2}) by applying the second rule of (\ref{qmaps}), $\Dk=(\KK-\KK_q)/\KK_q$, while $J_q$ is the q--scalar associated (via (\ref{qmaps})) to the energy density flux defined from the (or defining a) local $J$.

Notice that (\ref{cHam2}) is the quasi--local Hamiltonian constraint, which (from (\ref{HKq})) takes the functional form of an FLRW Friedman equation. Also, the evolution equations (\ref{evHH_q2}--\ref{eve_q2}) for the q--scalars $\HH_q, \rhoqe, \rhoqm$ are formally equivalent to FLRW equations for a CDM-DE mixture with EOS (\ref{eoscdm}) and (\ref{eosde}). This reinforces the interpretation of the q--scalars as averaged LTB scalars that mimic at every comoving shell $r=r_i$ the corresponding scalars of an FLRW background metric. In fact, as shown in \cite{RadAs}, an asymptotic FLRW background follows as all fluctuations $\delta^m,\,\delta^e\,\Dh,\,\Dj$ vanish in the limit $r\to\infty$ for all $t$, which is equivalent to the fact that in this limit the full system above reduces to the FLRW evolution equations (\ref{evHH_q2}--\ref{eve_q2}) and the Friedman equation in (\ref{cHam2}).

The autonomous system (\ref{evHH_q2}--\ref{evDh_q2}) can be solved numerically for a choice of $w$ and $J_q$, which determines $\Dj$ through (\ref{qmaps}). In the present work we will consider an interaction term $J_q$ proportional to the DE density and Hubble q--scalars as follows:
\be
J_{q} = 3\,\alpha\,\HH_{q}\,\rhoqe\quad\Rightarrow\quad \Dj=\Dh+\De,\label{intj_e}
\ee
where $\alpha$ is an dimensionless coupling constant. Note that the interaction energy flows from DE to CDM for $\alpha>0$ and  from  CDM to DE for $\alpha<0$.

Interactive mixtures with the EOS (\ref{eoscdm}) and (\ref{eosde}) and the interaction energy flux term  (\ref{intj_e}) were considered for an FLRW cosmology in \cite{copeland, ol, Maar}. This type of FLRW models provide background for a first order gauge invariant perturbation treatment that yields linear evolution equations for the associated perturbations of all sources, including the interaction term $J$, which can be considered as a phenomenological ``black box'' or (ideally in principle) related to some (yet unknown) early Universe physics.

As shown in \cite{suss15}, the dynamics of LTB metrics in the q--scalar formalism yields (through evolution equations like (\ref{evHH_q2}--\ref{evDh_q2})) an exact non--linear generalisation of linear gauge invariant cosmological perturbations in the isochronous gauge (for any source compatible with the LTB metric). The advantage of using numerical solutions of  (\ref{evHH_q2}--\ref{evDh_q2}) lies in the possibility to examine in the non--linear regime the connection between the assumptions on the CDM--DE interaction mediated by $J$ and observations on structure formation in the galactic and galactic cluster and supercluster scales.

\section{The dynamical system}\label{dynsys}

The q--scalar formalism described in the previous section allows us to define suitable dimensionless functions that transform the system (\ref{evHH_q2}--\ref{evDh_q2}) into an autonomous five-dimensional dynamical system that is amenable to a qualitative phase space analysis, analogous to that undertaken in \cite{izsuss17}.

For the density scalar functions we define below the following q--scalars that are analogous to the $\Omega$ factors in FLRW cosmologies
\bse\ba
\Ommq=\frac{\kappa\rhoqm}{3\HH_q^2}, \qquad \frac{\dot\Ommq}{\HH_q}=\frac{\kappa \dot\rhoqm}{3\HH_q^3}-\frac{\Ommq\dot{\HH}_q}{\HH_q},\label{Ommq}\\
\Omeq=\frac{\kappa\rhoqe}{3\HH_q^2},\qquad \frac{\dot\Omeq}{\HH_q}=\frac{\kappa \dot\rhoqe}{3\HH_q^3}-\frac{\Omeq\dot{\HH}_q}{\HH_q},\label{Omeq}
\ea\ese
transforming the Hamiltonian constraints in (\ref{cHam2})--(\ref{cHam2b}) into the following elegant forms
\ba
\Ommq+\Omeq-1=\Omkq,\qquad \Omkq=\frac{\KK_q}{\HH^2_q},\label{HamC1} \\ 2\Dh = \Ommq\,\Dm + \Omeq\,\De-\Omkq\,\Dk.\label{HamC2}
\ea
Next, we introduce a dimensionless coordinate $\xi(t,r)$ that will serve as the phase space evolution parameter, so that for all comoving curves $r=r_i$ we have
\begin{equation}\frac{\partial}{\partial\xi}=\frac{1}{\HH_q}\frac{\partial}{\partial t}.\label{xidef}\end{equation}
In terms of $\xi$ and using the interaction term defined in (\ref{intj_e}), the system (\ref{evHH_q2}--\ref{evDh_q2}) is transformed into the following dynamical system
\bse\label{eveqs_q3}
\ba
\frac{\partial{\Ommq}}{\partial{\xi}} &=& \Ommq\,\left[ -1+\Ommq+\left( 1+3\,w\right)\,\Omeq\right]+3\,\alpha \Omeq, \label{sistdinint2a}\\
\frac{\partial{\Omeq}}{\partial{\xi}} &=& \Omeq\,\left[ \left( 1+3\,w \right) \left( -1+\Omeq \right)+\Ommq -3\,\alpha\,\right], \label{sistdinint2b}\\
\frac{\partial{\Dm}}{\partial{\xi}} &=& -3\,\Dh\,\left(1+\Dm \right)+3\alpha \frac{\Omeq}{\Ommq}\left(\De+\Dh-\Dm\right), \label{sistdinint2c}\\
\frac{\partial{\De}}{\partial{\xi}} &=&-3\Dh\left(1+w+\De+\alpha \right), \label{sistdinint2d}\\
\frac{\partial{\Dh}}{\partial{\xi}} &=& -\Dh\left( 1+3\Dh \right)+\frac{\Ommq\,\left( \Dh-\Dm \right)}{2}\nonumber\\
&&\qquad \qquad+\frac{\left( 1+3\,w \right)\,\Omeq\,\left(\Dh-\De \right)}{2}. \label{sistdinint2e}
\ea
\ese
The system (\ref{sistdinint2a}-\ref{sistdinint2e}) can be solved numerically for a set of initial conditions for every comoving shell $r=r_i$ once we fix the free parameters of the model, $w$ and $\alpha$. We can compute afterwards $\HH_q(\xi,r_i)$ from
\be
\frac{\partial{\HH_q}}{\partial{\xi}}=\frac{\dot{\HH}_q}{\HH_q}=-\HH_q\left(1+\frac{1}{2}\Ommq+\frac{1+3w}{2}\Omeq\right).\label{HOm}
\ee
Once we have computed the phase space variables $\Ommq,\,\Omeq,\,\Dm,\,\De,\,\Dh$, all relevant quantities can be obtained: The q--scalars associated with the CDM and DE densities and the spatial curvature and its fluctuation $\Dk$ follow directly from (\ref{Ommq})--(\ref{Omeq}), (\ref{HamC1}), (\ref{HamC2})  and (\ref{HOm}), while all local quantities follow from $A=A_q(1+\delta^A)$ for $A=\HH,\,\KK,\,\rhom,\,\rhoe,\,J$.

Additionally, it is possible to recover physical time from the phase space evolution parameter $\xi(t ,r)$ from evaluating at each fixed $r=r_i$,
\be
t(r_i)= \int_0^{\xi(t,r_i)} {\frac{d \xi'}{\HH_q(\xi',r_i)}},\label{phystime}
\ee
though it is important to bear in mind that hypersurfaces of constant $\xi$ and $t$ do not coincide, thus for every scalar $A$ we have $[\partial A/\partial r]_t\ne [\partial A/\partial r]_\xi$ (the appropriate integrability conditions are discussed in detail in \cite{izsuss10}). Taking this into account, the LTB metric functions follow from evaluating $R=\exp\left(\int{\HH_q dt}\right)$ and $R'=R\exp\left(-\int{\HH_q \Dh dt}\right)$ for $\HH_q,\,\Dh$ as functions of $(t,r)$.

\subsection{Homogeneous and inhomogeneous subspaces}

As in \cite{izsuss17,izsuss10}, we can split the phase space of (\ref{sistdinint2a}-\ref{sistdinint2e}) into two interrelated projection subspaces:
\begin{description}
\item[The homogeneous subspace.] It is defined by the phase space variables $\Ommq$ and $\Omeq$, since they are fully determined by the evolution equations (\ref{sistdinint2a})--(\ref{sistdinint2b}), which do not involve the other phase space variables $\Dm,\,\De,\,\Dh$. In fact, for every trajectory (fixed $r$) these evolution equations are formally identical to FLRW equations for the analogous variables.
\item[The inhomogeneous subspace.] It is defined by the remaining three phase space variables $\Dm,\,\De,\,\Dh$, as these provide a measure of the departure of the local scalars from their homogeneous FLRW counterparts.
\end{description}
The study of the phase space will be undertaken by looking at its trajectories in terms of these two projections.
\begin{table}[]
\centering
\caption{The critical points and their respective eigenvalues of the system (\ref{sistdinint2a}-\ref{sistdinint2e}).}
\label{criticalpoints}
\begin{tabular}{|l|l|l|}
\hline
\begin{tabular}[c]{@{}l@{}}Critical\\  points\end{tabular} & $\left(\Ommq,\Omeq,\Dm,\De,\Dh\right)$ & Eigenvalues \\ \hline
PC1 & $\left(1,\,0,\,0, \De \hbox{arbitrary},\,0\right)$ & \begin{tabular}[c]{@{}l@{}} $\lambda_1 = 0$,\,$\lambda_2 = -\frac{3}{2}$,\,$\lambda_3 = -3\,(w+\alpha)
$,\,$\lambda_4 = \lambda_5 = 1.$\end{tabular} \\ \hline
PC2 & $\left(1,\,0,\,-1,\,-(1+w+\alpha)
,\,-\frac{1}{2}\right)$ & \begin{tabular}[c]{@{}l@{}}$ \lambda_1= -3\,(w+\,\alpha),\,\lambda_2 = \lambda_3 = \frac{3}{2},\,\lambda_4 = \frac{5}{2},\,\lambda_5 = 1.$\end{tabular} \\ \hline
PC3 & $\left(1,\,0,\, -1,\, -(1+w+\alpha),\,\frac{1}{3}\right)$ & \begin{tabular}[c]{@{}l@{}}$\lambda_1 = -3\,(w+\,\alpha),\,\lambda_2 = -\frac{5}{2},\,\lambda_3 = 1,\,\lambda_4 =\lambda_5 = -1. $\end{tabular} \\ \hline
PC4 & $\left(-\frac{\alpha}{w},\,1+\frac{\alpha}{w},\, 0,\,0,\,0\right)$ & \begin{tabular}[c]{@{}l@{}}$\lambda_1 =\lambda_2 = 1+3(w+\alpha),\,\lambda_3 = \lambda_4 = 3(w+\alpha)$,\\$
\,\lambda_5 = -\frac{3}{2}\left( 1+w+\alpha \right)\,.$\end{tabular} \\ \hline

PC5 & \begin{tabular}[c]{@{}l@{}}$\left(-\frac{\alpha}{w},\,1+\frac{\alpha}{w},-\frac{4\,w\,\alpha+3\,w^2+\alpha^2+\alpha+w}{\alpha},\right.$\\ \qquad\qquad $\left. -(1+w+\alpha),\,w+\alpha\right)$ \end{tabular}& \begin{tabular}[c]{@{}l@{}}$ \lambda_1 = 1,\,\lambda_2 = 1+3(\,\alpha+\,w),\,\lambda_3 = -\frac{3}{2}\left(1+3(w+\alpha)\right),$\\ $\,\lambda_4 = 3(w+\alpha),\,\lambda_5 = -3(w+\alpha)
$\end{tabular} \\ \hline

PC6 & \begin{tabular}[c]{@{}l@{}}$\left(-\frac{\alpha}{w},\,1+\frac{\alpha}{w},\,-(1+w+\alpha),\,\right.$ \\ \qquad$\left.-(1+w+\alpha),\,-\frac{1}{2}(1+w+\alpha)\right)$\end{tabular} & \begin{tabular}[c]{@{}l@{}}$\lambda_1 =1+3\,(\alpha+\,w),\,\lambda_2 = \frac{3}{2}(1+w+ \,\alpha),\,$\\ $\lambda_3 = \frac{5+9\,(w+\alpha)}{2},\,$\\ $\lambda_4 = \frac{3}{2}\left(1+3(w+\alpha)\right),\,\lambda_5 = 3\,(w+\alpha)
.$\end{tabular} \\ \hline

PC7 & \begin{tabular}[c]{@{}l@{}}$\left(-\frac{\alpha}{w},\,1+\frac{\alpha}{w},\, -(1+w+\alpha),\,\right.$\\ $\qquad\left. -(1+w+\alpha), \frac{1}{3}+\alpha+w\right)$\end{tabular} & \begin{tabular}[c]{@{}l@{}}$ \lambda_1 = -1-3\,(w+\,\alpha),\,\lambda_2 = -1,\,$\\ $\lambda_3 = 1+3\,(w+\alpha),\,\lambda_4 =-\frac{5+9\,(w+\alpha)}{2},\lambda_5 =3(w+\alpha)
$\end{tabular} \\ \hline
\end{tabular}
\end{table}

\subsection{Critical points}

The critical points of the system (\ref{sistdinint2a}-\ref{sistdinint2e}) and their respective eigenvalues are shown in table \ref{criticalpoints}. As expected, they depend on the free parameters $w$ and $\alpha$, save for $PC1$. The critical point $PC1$ is in fact a line parallel to the $\De$ axis. The eigenvalue $\lambda_1$ of $PC1$ is zero, corresponding to a eigenvector that is also parallel to the $\De$ axis, indicating that near the line there is no evolution of the space phase trajectory in that direction. For $\alpha<0$, the critical points $PC4$, $PC5$, $PC6$ and $PC7$ are non physical as their component $\Ommq$ is negative, which means the CDM energy density should be negative. We will examine below the homogeneous subspace closely.

\subsection{Homogeneous subspace.}

The homogeneous subsystem for $\alpha>0$ has the following critical points (see figure \ref{fig1}):
\begin{itemize}
\item  future attractor:  $PCA=[\Ommq,\Omeq]_{\hbox{\tiny{PCA}}}=[-\alpha/w,\,\,1+\alpha/w]$
\item  past attractor:    $PCR=[\Ommq,\Omeq]_{\hbox{\tiny{PCR}}}=[1,\,\,0]$
\item  saddle point:    $PCS=[\Ommq,\Omeq]_{\hbox{\tiny{PCS}}}=[0,\,\,0].$
\end{itemize}
Both, $PCA$ and $PCR$ can be considered as critical points of the phase space that would result from an FLRW model, or as a projection of the $PC1-PC7$ points over the $[\Ommq,\Omeq]$ subspace in a full five-dimensional representation. In the former case, the trajectories in the phase-space are computed for a given set of initial conditions with $\Dm=\De=\Dh=0$ and live completely in the homogeneous space, while in the later case the trajectories are computed with a general choice of $\Dm,\,\De$ and $\Dh$ and are represented in the homogeneous subspace as projections of the five--dimensional space trajectories over the $[\Ommq,\Omeq]$ subspace. Additionally, in a similar way as in the interaction used in \cite{izsuss17}, we have a one--dimensional invariant subspace (a line) given in this case by
\be
\Omeq=-\frac{w+\alpha}{\alpha}\Ommq\quad \Rightarrow\quad \frac{\partial}{\partial \xi}\left(\Ome+\frac{w+\alpha}{\alpha}\Omm\right)=0, \label{invline}
\ee
where we used (\ref{sistdinint2a}-\ref{sistdinint2b}). This invariant line contains both the saddle point and the future attractor. Hence,  the system can evolve from the saddle point to the future attractor (for initial conditions with ${\Ommq}(0)<-\alpha/w$), or from past infinity ($\xi\to-\infty$) to the future attractor (for $\Ommq(0)>-\alpha/w$). The $[\Ommq,\Omeq]$ plane is divided in two regions by this invariant line: the region where trajectories evolve from the $\Ommq=0$ axis to the future attractor and the region where the trajectories evolve from the past attractor. The later region contains part of the attraction basin of $PCA$: trajectories that evolve from the past attractor to the future attractor, representing an ever expanding scenario where initially there is only CDE with CDM density increasing from the interaction with DE. This region also contains trajectories for which $\Ommq,\,\Omeq\to\infty$, which correspond to comoving layers that bounce (since $\Ommq,\,\Omeq$ diverge as $\HH_q\to 0$). We will not consider the evolution of such trajectories.

For $\alpha<0$ the future attractor $PCA$ lies in an unphysical phase space region marked by negative $\Ommq$. For trajectories emerging  from the past attractor the physical evolution, which can only be defined up to the invariant line, describes an expanding scenario in which energy density flows from the CDM to the DE component until the CDM density vanishes on the comoving shells (at different times for different shells). However, the fact that the past evolution is not unphysical makes the coupling term (\ref{intj_e}) acceptable also when $\alpha<0$, as has been stated in the literature \cite{GZun14} deling with these CDM-DE mixtures in FLRW cosmologies. This stands in sharp contrast with the coupling used in \cite{izsuss17}, where $\alpha<0$ leads to grossly unphysical past evolution, which implies considering only the coupling with $\alpha>0$ (as in FLRW cosmology scenarios).

\subsection{Initial conditions, scaling laws and singularities.}

To specify initial conditions to integrate the dynamical system (\ref{sistdinint2a})--(\ref{sistdinint2e}) we need to provide an initial value formulation for the LTB models under consideration. Proceeding as in \cite{izsuss17}, we specify initial conditions given at an arbitrary hypersurface $t=t_0$ (subindex ${}_0$ will denote henceforth evaluation at $t=t_0$). It is useful to write LTB metric (\ref{LTB}) in the following FLRW--like form
\ba \dd s^2 = -\dd t^2 +L^2\,\left[\frac{\Gamma^2\,R'_0{}^2 \dd r^2}{1-\KK_{q0} R_{0}^2 }+R_{0}^2\,(\dd\theta^2+\sin^2\theta\dd\phi^2)\right],\label{LTB2}\\
\Gamma = 1+ \frac{L'/L}{R'_0/R_0},\ea
where $L=L(t,r)$ is analogous to the FLRW scale factor. Since the LTB metric admit an arbitrary rescaling of the radial coordinate, we can always define a convenient radial coordinate by specific choices of $R_0(r)$. We can identify $L=0$ as the locus of the Big Bang singularity, while $\Gamma=0$ marks the locus of a shell crossing singularity \cite{izsuss10}.

From (\ref{HKq}), (\ref{evHH_q2}), (\ref{evm_q2}) and (\ref{eve_q2}) we can see how the q--scalars scale as their equivalent FLRW scalars, that is
\ba
\fl \KK_q=\frac{\KK_{q0}}{L^2},\quad
\rhoqm =\frac{\rhoqmi}{L^3}\left[1+\frac{\alpha}{w+\alpha}\left(1-\frac{1}{L^{1+w+\alpha}}\right)\right],\quad
\rhoqe =\frac{\rhoqei}{L^{3(1+w+\alpha)}},\label{scalinglawrhomq}
\ea
which lead to
\be\fl \HH_q^2=\left(\frac{\dot{L}}{L}\right)^2 = \frac{\kappa}{3 L^3}\left\{\rhoqmi\left[1+\frac{\alpha}{w+\alpha}\left(1-\frac{1}{L^{1+w+\alpha}}\right)\right]+\frac{\rhoqei}{L^{1+w+\alpha}}\right\}-\frac{\KK_{q0}}{L^2},\label{EvL}\ee
Initial conditions to integrate the system (\ref{sistdinint2a}-\ref{sistdinint2e}) follow from specifying initial profiles $\rhomi(r), \rhoei(r)$ and $\KK_{0}(r)$ and a given choice of $R_0(r)$.  The initial profiles of the q--scalars $\rhoqmi(r),\,\rhoqei(r),,\,\KK_{q0}(r)$ and the fluctuations $\Dm_0,\,\De_0,\,\Dh_0$ follow directly from (\ref{qmaps}) with $R=R_0$. These initial profiles must comply with the condition $\Dm_0,\,\De_0,\,\Dh_0\to 0$ as $R_0(r)\to\infty$ to guarantee the existence of an asymptotic FLRW background (see \cite{RadAs}). 

For simplicity, $\xi$ can take as initial value  $\xi_0=\xi(t_0,r)=0$ for all $r$, which sets the initial conditions for the dynamical system:  $\Ommqi=\Ommq(0,r)$, $\Omeqi=\Omeq(0,r)$, $\Dm_0=\Dm(0,r)$, $\De_0=\De(0,r)$ and $\Dh_0=\Dh(0,r)$. This choice of initial value of $\xi$ means that $\xi=0$ and $t=t_0$ mark the same hypersurface, though hypersurfaces of constant $t$ and $\xi$ are different for $\xi\ne 0$ and $t\ne t_0$.

Whether comoving shells expand or bounce/recollapse can be determined from (\ref{EvL}) by looking at the roots of $\HH_q$ through ${\dot{L}}^2=L^{-1}Q(L)$ where
\ba
\fl Q(L) = L^3\HH_q^2 =\HH_{q0}^2\left[a+\frac{b}{L^{3(w+\alpha)}}+c L\right]=\HH_{q0}^2\left[a+b\,\hbox{e}^{-3(w+\alpha)\xi}+c\,\hbox{e}^\xi\right],\\
\fl\hbox{with:}\quad  a=\Ommqi+\frac{\alpha\,\Omeqi}{w+\alpha},\qquad b=\frac{w\,\Omeqi}{w+\alpha},\qquad c = 1-\Omeqi-\Ommqi=\Omkqi.\ea
where we used the fact that $\dd\xi=\HH_q \dd t$ leads to $\xi=\ln(L)$. For each choice of initial conditions $\Ommqi,\,\Omeqi$ and free parameters $w$ and $\alpha$, if  $Q(L)$ has real roots for a given comoving shell, the latter will bounce at the value of $\xi$ where $Q=0$. Conversely, if $Q(L)$ has no real roots the layer has an ever expanding evolution. For the remaining of the paper we will only consider the phase space evolution of expanding layers.

To obtain analytic solutions we need to solve (\ref{EvL}) and obtain $\Gamma$ from the relation $L'/L=R'_0(\Gamma-1)/R_0$, where the radial derivatives must be evaluated for constant $t$ (see \cite{izsuss17})). The scaling laws for the fluctuations can be found from (\ref{scalinglawrhomq}). For example, using the definition of $\De$, it is straightforward to show that
\be
\fl\De=-(1+w+\alpha)+\frac{1+w+\alpha+\De_0}{\Gamma}\quad \Rightarrow \quad \Gamma=\frac{1+w+\alpha+\De_0}{1+w+\alpha+\De},\label{Deslaw}
\ee
which can (in principle) be used to evaluate $\Gamma$ once we have a solution $t=t(L,r)$ of (\ref{EvL}). Since analytic solutions of (\ref{EvL}) may only exist for very restricted values of $\alpha,w$, scaling laws like (\ref{Deslaw}) are not useful. In general, the evolution of the models needs to be determined numerically.

If $\Gamma=0$ we have a shell crossing singularity, which means that initial conditions should be found to avoid this happening (it is not possible to provide simple guidelines for this, as in dust solutions with zero cosmological constant, see \cite{ltbrev1,ltbrev2,ltbrev3,RadProfs}). As shown in previous work (for example \cite{izsuss17}) the q--scalar formalism fails at shell crossing singularities because the fluctuations $\Dm,\,\De,\,\Dh$ diverge. Hence, we will select initial conditions such that shell crossings are avoided: $\Gamma>0$ holds throughout the full phase space evolution.

\section{Critical points in terms of the parameters $w$ and $\alpha$}\label{criticalpointsLTB}

In this section the critical points of the system are studied for different possibilities of the free parameters $w$ and $\alpha$. As both parameters are widely used in the FLRW model, we will consider a parameter range that is common in Cosmology.

The constant EOS parameter $w$ plays a similar role as in analogous CDM-DE mixtures based on FLRW models. While observational data seems to favor  the $\Lambda$--CDM model for which $w=-1$ holds exactly, small variations from this value are still possible \cite{plank}. We will henceforth adopt the current terminology, in the literature by referring to DE with $w>-1$ as ``quintessence models'' and $w<-1$ as ``phantom models''. The latter models present several theoretical problems, such as the violation of the second law of thermodynamics once we assign an entropy to the phantom fluid, or the presence of a negative kinetic energy of the phantom field term (when described by a scalar field) \cite{copeland}. In this article we will consider quintessence and phantom models with $w$ close to $-1$.

Considering the same interaction in the present article, the CDM--DE mixture in FLRW geometry examined in \cite{db} finds that the second law of thermodynamics (based on the entropy of DE as an effective field) is violated if $\alpha<0$, while the entropy is zero for a scalar field in a pure quantum state. They also find that for positive $\alpha$ it is necessary to impose the bound $\alpha<0.1$ in order to reproduce the observed values of BAO and CMB anisotropy \cite{ol,ol2}. On the other hand, in \cite{smlm} the evolution of  linear perturbations in a FLRW background sharing our assumptions on CDM, DE and $J_q$ leads to the bounds $-0.22>3\alpha>-0.90$ to comply with the constraints of CMB anisotropy.  These results are specially interesting, since the dynamics of LTB solutions described by q--scalars and their fluctuations can be mapped to linear perturbation on an FLRW background \cite{suss15}. While the spherical symmetry of LTB models allows for the description of a single structure, the latter can be studied exactly in full non--linear regime. In the present paper we will assume positive and negative values of $\alpha$.

The critical points $PC4-PC7$ share the values $\Ommq=-\alpha/w$ and $\Omeq=1-\alpha/w$ of the homogeneous projection, but are distinct in the inhomogeneous projection coordinates. The points $PC5$ and $PC7$ are always saddle points in the range of free parameters considered. For a choice of parameters such that $1+w+\alpha>0$, the critical point $PC4$  is a future attractor as all their eigenvalues are negative defined, while $PC6$ is saddle point. On the other hand, when $1+w+\alpha=0$, $PC6$ has the same components as $PC4$, and behaves as a non hyperbolic point as one of its eigenvalues are null while the rest are negative defined. Finally, for $1+w+\alpha<0$, $PC6$ is the future attractor while $PC4$ is a saddle point.

\begin{figure}[tbp]
\includegraphics*[scale=0.30]{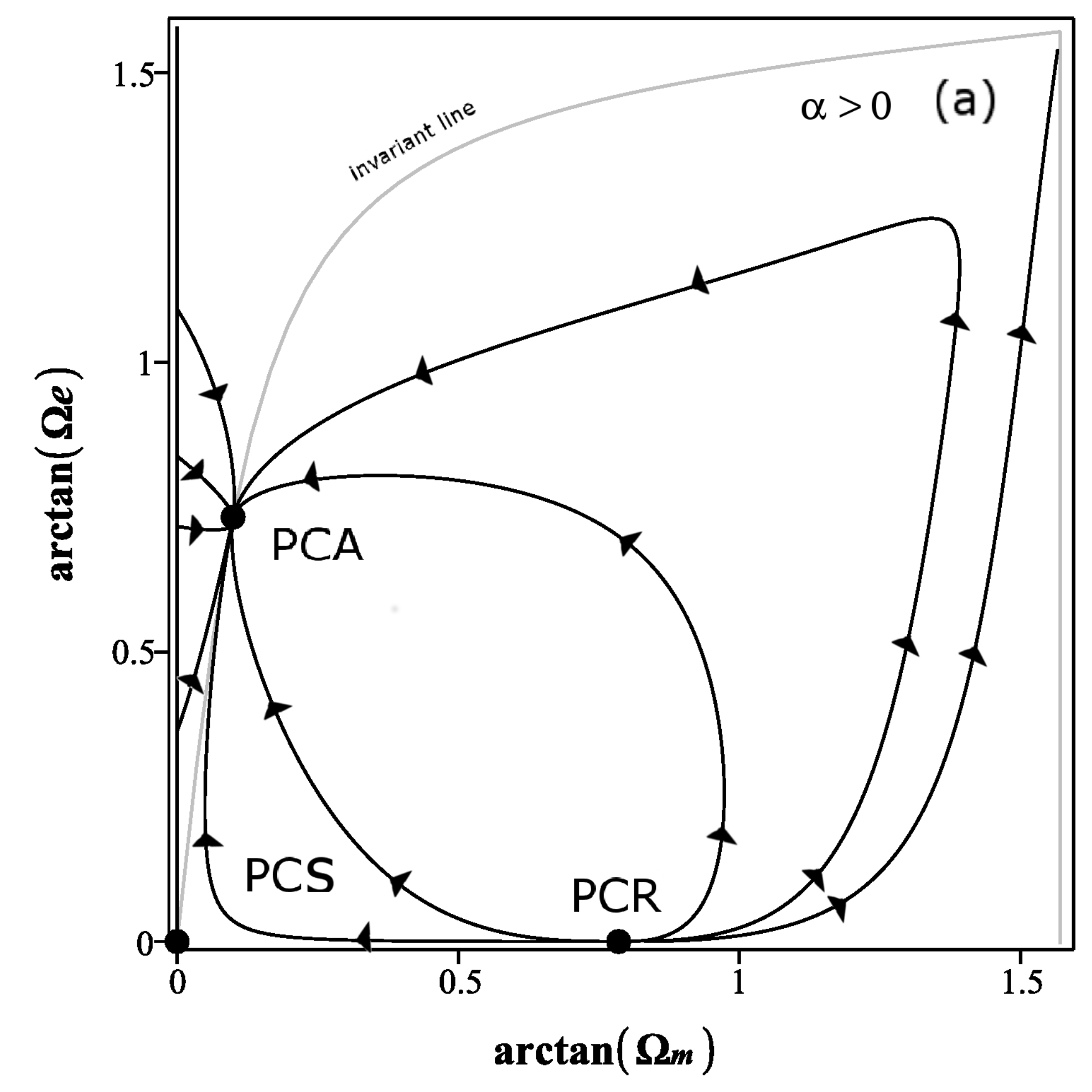}
\includegraphics*[scale=0.30]{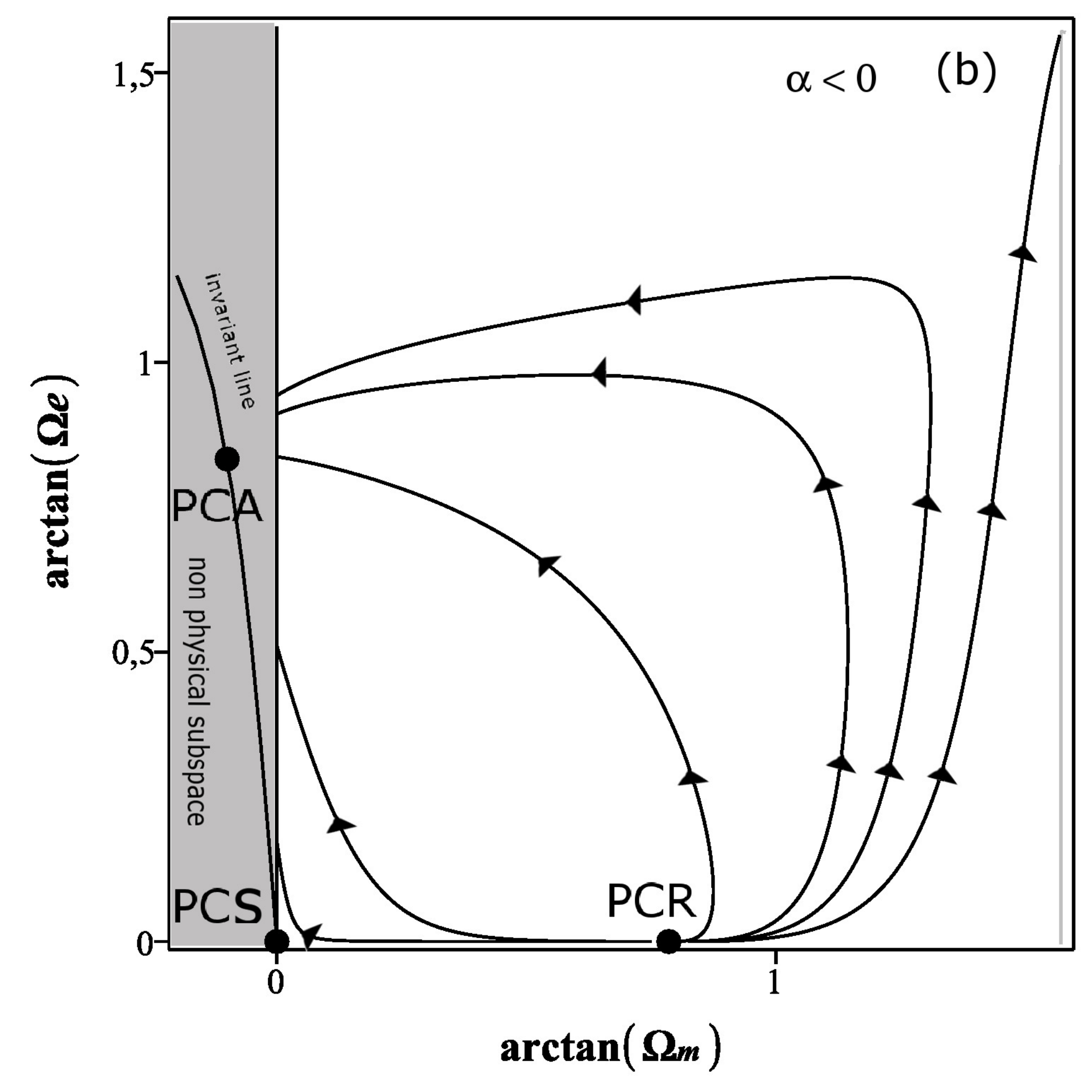}
\caption{Panel (1a): Critical points and numerical trajectories of the dynamical system (\ref{sistdinint2a}-\ref{sistdinint2e}) in the homogeneous projection for $\alpha=0.1$ and $w=-1.0$. For other choices of the parameters with $\alpha>0$ the point $PCA$ will be in a different position, and, consequently, the invariant line will have a different slope. For some initial conditions choice, the trajectory evolves to the future attractor $PCA$ from the $\Ommq=0$ axis or from the past attractor $PCA$, or  it diverges. Panel (1b): Critical points and numerical trajectories of the dynamical system (\ref{sistdinint2a}-\ref{sistdinint2e}) in the homogeneous projection for $\alpha=-0.1$ and $w=-1.0$. For some initial conditions choice the trajectory evolves to $\Ommq=0$ axis in the future, or it diverges.} \label{fig1}\end{figure}

Figure \ref{fig1} shows the homogeneous subspace together with the critical points $PCR$, $PCA$,  $PCS$ and the invariant line for both cases: $\alpha>0$ in panel (a) (in this case we have chosen $\alpha=0.1$ and $w=-1$), and $\alpha<0$ in panel (b)($\alpha=-0.1$ and $w=-1$). Some numerically computed trajectories are shown for illustration purposes only. We have chosen to represent $\arctan(\Ommq)\,\, \hbox{vs.}\,\,\arctan{\Omeq}$ in order to deal with finite values in the plots. The same criterion is used for the rest of the homogeneous projection plots.

\begin{figure}[tbp]
\includegraphics*[scale=0.30]{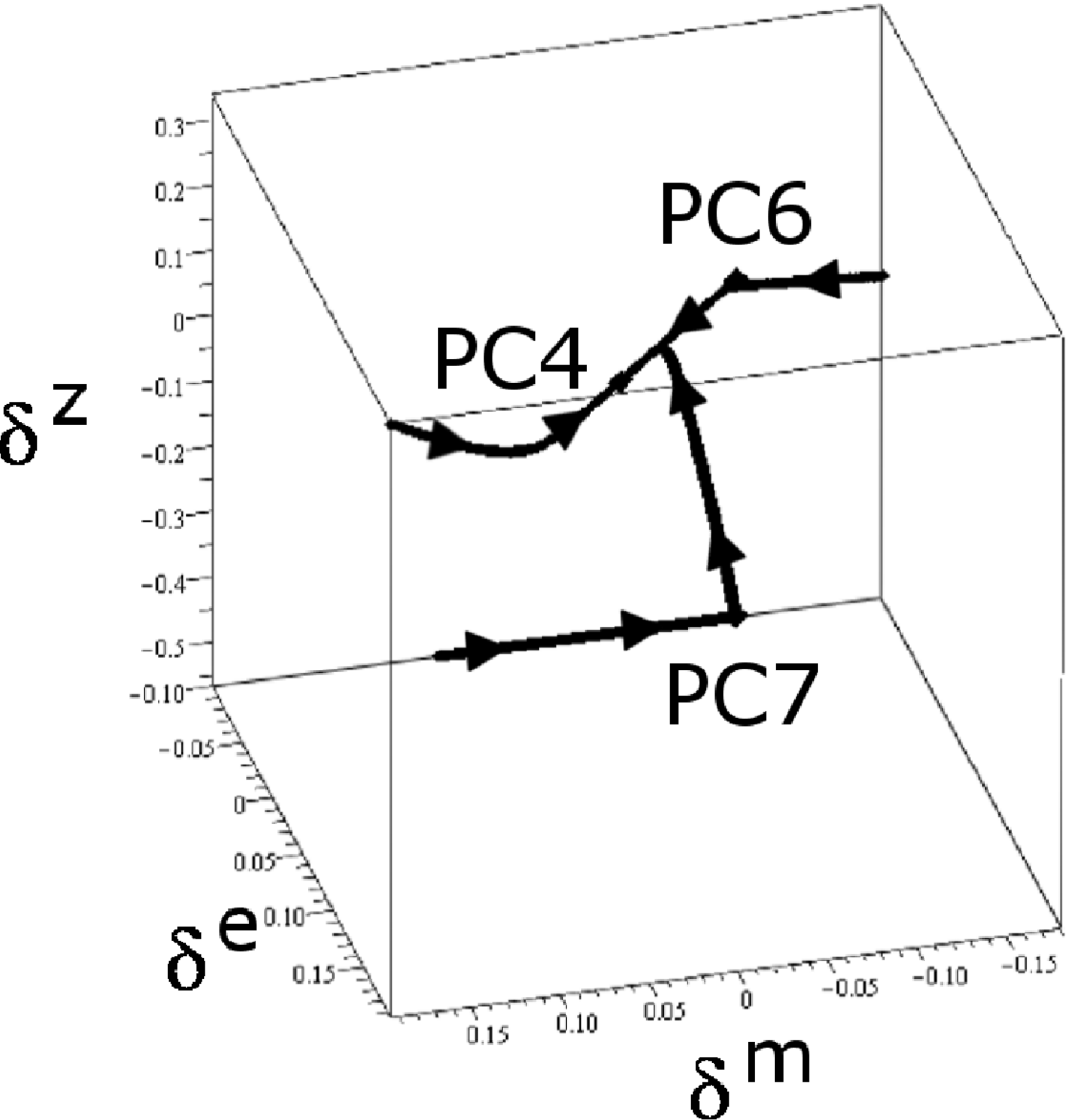}
\includegraphics*[scale=0.3]{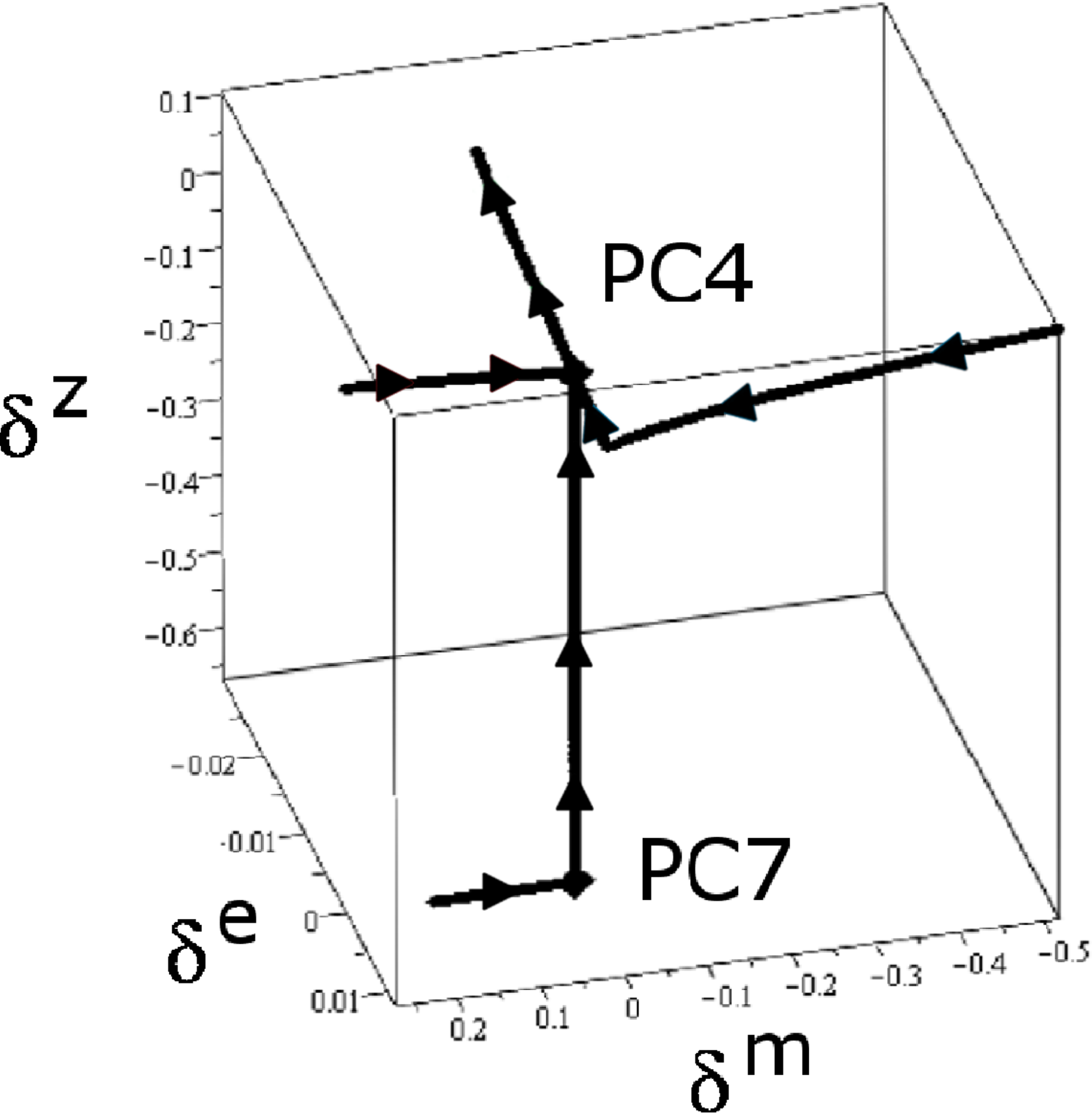}
\includegraphics*[scale=0.3]{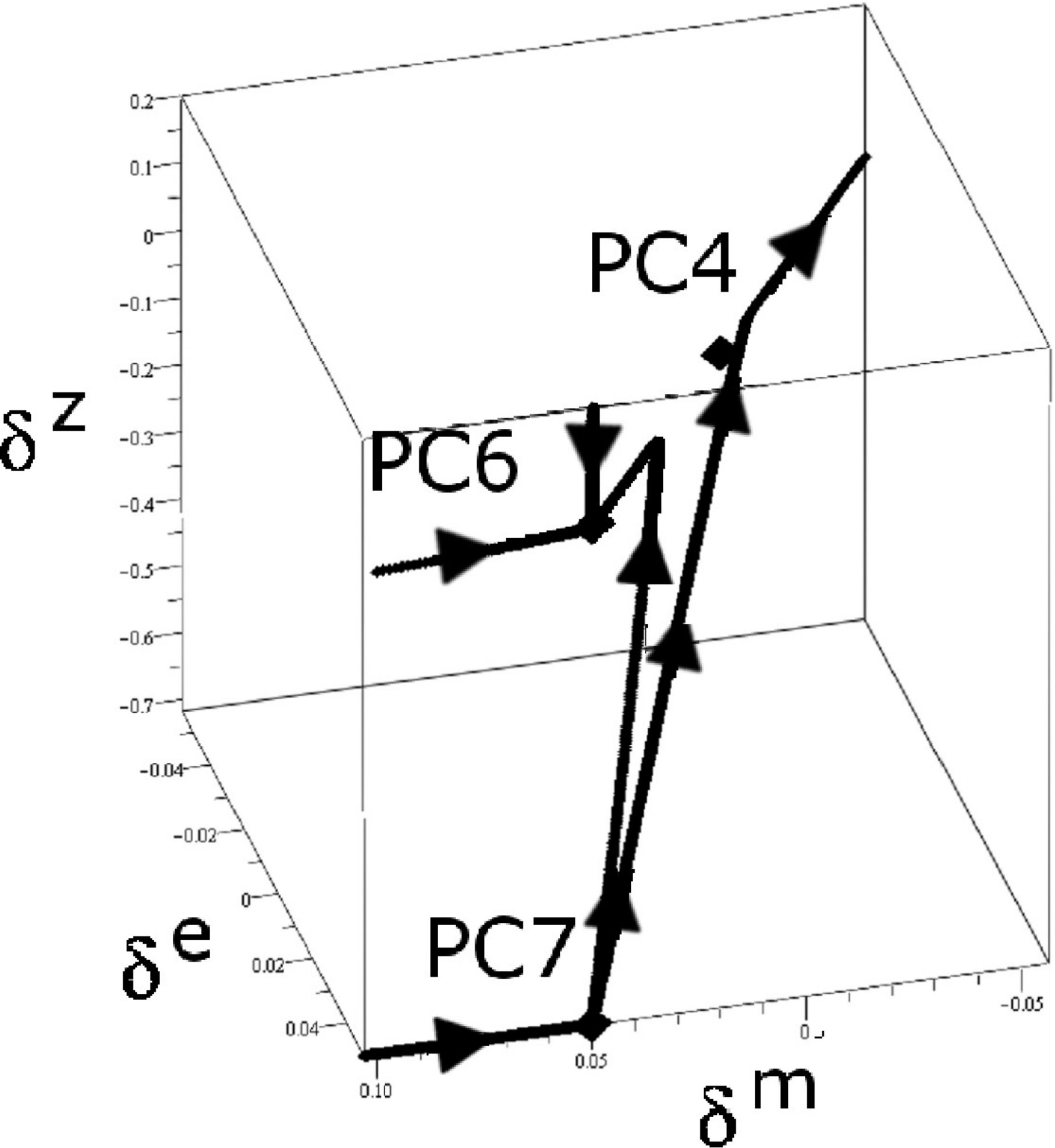}
\includegraphics*[scale=0.3]{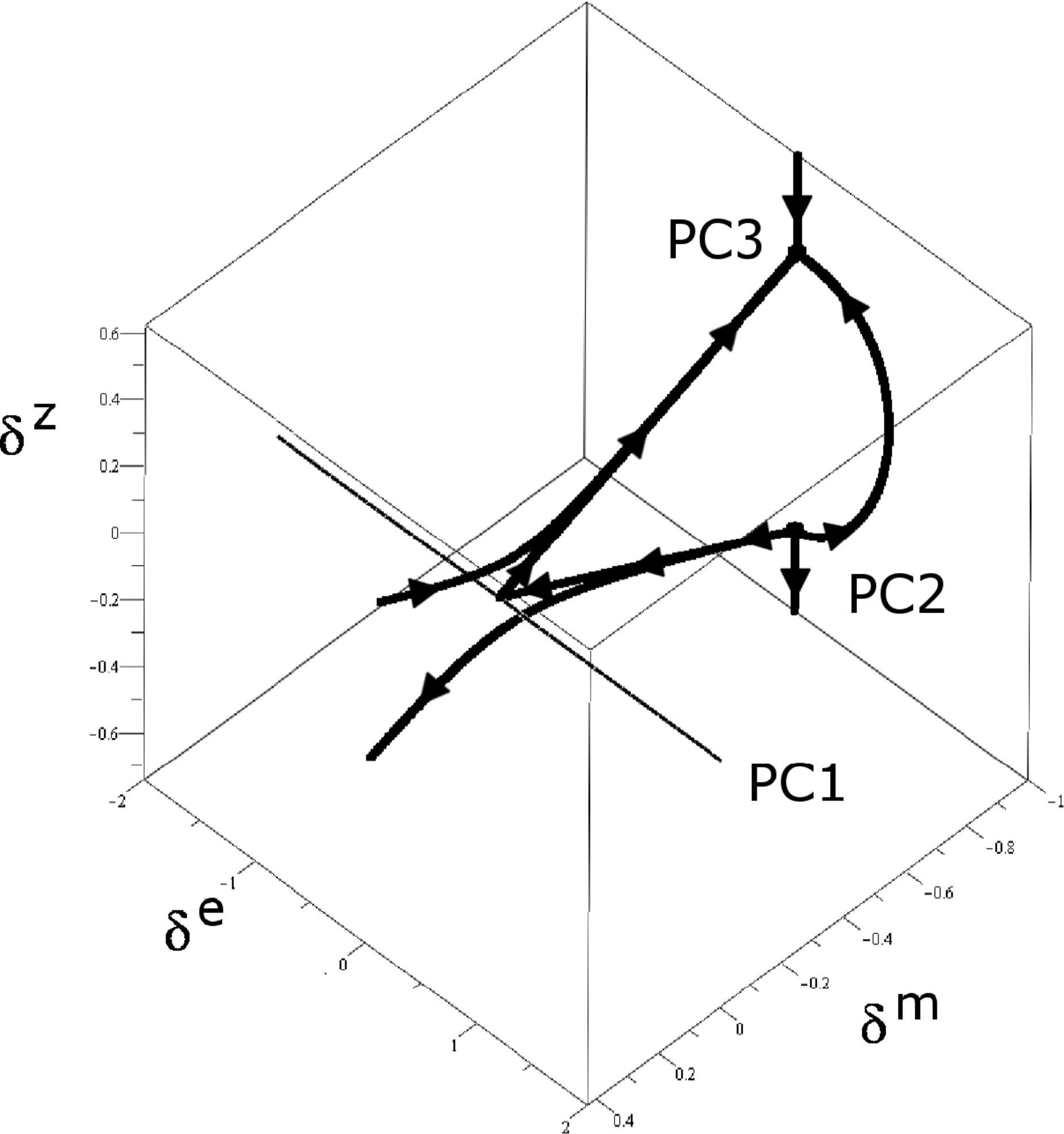}
\caption{Critical points and numerical trajectories of the dynamical system (\ref{sistdinint2a}-\ref{sistdinint2e}) in the inhomogeneous projections. Panel (2a): Inhomogeneous subspace $\Dm$ vs. $\De$ vs. $\Dh$ with $\Ommq=-\alpha/w$ and $\Omeq=1-\alpha/w$ for $\alpha=0.1$ and $w=-0.9$. Panel (2b): Inhomogeneous subspace $\Dm$ vs. $\De$ vs. $\Dh$ with $\Ommq=-\alpha/w$ and $\Omeq=1-\alpha/w$ for $\alpha=0.1$ and $w=-1.0$. Panel (2c): Inhomogeneous subspace $\Dm$ vs. $\De$ vs. $\Dh$ with $\Ommq=-\alpha/w$ and $\Omeq=1-\alpha/w$ for $\alpha=0.1$ and $w=-1.15$. Panel (2d): Inhomogeneous subspace  $\Dm$ vs. $\De$ vs. $\Dh$ with $\Ommq=1$ y $\Omeq=0$ for $\alpha=0.1$ and $w=-1.0$.} \label{fig2}\end{figure}

Figure \ref{fig2} shows the two inhomogeneous projections of the system (\ref{sistdinint2a}-\ref{sistdinint2e}) and some numerically computed trajectories. Panel (a), (b) and (c) represent the projection with $\Ommq=-\alpha/w$ and $\Omeq=1-\alpha/w$ and the critical points $PC4$, $PC6$ and $PC7$ for different choices of $w$ and $\alpha>0$: panel (a) shows a set of parameters where $1+w+\alpha>0$ and the future attractor is $PC4$; panel (b) shows a set with $1+w+\alpha=0$ where $PC4$ and $PC6$ are the same point; and, finally, panel (c) shows a set of parameters is represented where $1+w+\alpha<0$ and the future attractor is $PC6$. The critical point $PC5$ is not represented as it is always a saddle point with a large $\Dm$ component, far away from the rest of points. Panel (d) shows the projection $\Ommq=1$ and $\Omeq=0$ for $\alpha=0.1$ and $w=-1$, although choosing a different value of $w$ and $\alpha$ will not change the general behavior of the points or the trajectories. The projection with $\Ommq=-\alpha/w$ and $\Omeq=1-\alpha/w$ will be unphysical when $\alpha<0$. The projection $\Ommq=1$ and $\Omeq=0$, on the other hand, will be physically plausible and phenomenologically identical to that in panel \ref{fig2}d in the $\alpha>0$ case.

\subsection{Energy density flow from DE to CDM ($\alpha>0$)\label{subsecalphapos}}
In this case all seven critical points are physical. The attractor is a different point for the different choices of $w$ and $\alpha$, as stated above. The presence of a future attractor for $\alpha>0$ allow us to find initial profiles that present inhomogeneities in the attraction basin of it, {\it i.e} the fluctuations ($\delta$ functions) evolving to constant values given by the components of the corresponding critical point. This behaviour is examined further ahead.

\subsubsection{Quintessence and cosmological constant cases}
When $w\geq-1$ and $\alpha>0$, the critical point $PC4$ acts as a future attractor and the trajectories nearby evolve to it. On the other hand, the critical point $PC2$ is a past attractor as all the eigenvalues of the system computed near $PC2$ have positive values. The rest of the critical points are saddle points with their own attraction subspace generated by the corresponding eigenvectors.

Panel \ref{fig1}a shows the homogeneous subspace for $w=-1$: it is formally identical to that of the $w>-1$ case, except for the position of the future attractor and the shape of the invariant line. In the panel \ref{fig2}a, the inhomogeneous projection $\Ommq=-\alpha/w$ and $\Omeq=1-\alpha/w$ is shown for the $w>-1$ case. The attractor $PC4$ is also shown together with some trajectories in its vicinity that evolve to it. Also, the saddle points $PC6$ and $PC7$ are displayed (the point $PC5$ is not shown given that it is located far away and is always a saddle point).  Finally, panel \ref{fig2}d displays the inhomogeneous subspace with $\Ommq=1,\,\Omeq=0$,  plotted for the $w=-1$ case but, again, it is phenomenologically identical to the $w>-1$ case. In this projection the past attractor $PC2$ and the saddle points $PC1,\, PC3$ are also displayed.

\subsubsection{Phantom dark energy case}
The points $PC4$ and $PC6$ can be (respectively) a future attractor and a saddle point when $1+w+\alpha>0$ or (respectively) a saddle point and a future attractor when $1+w+\alpha<0$. Finally, when $1+w+\alpha=0$ both points have the same coordinates and the point is non-hyperbolic, {\it i.e.} it is an attractor in some directions and there is no evolution near it in other directions. The rest of points behave in a similar way as in the previous cases for any choice of free parameters.

The homogeneous subspace is identical to that of panel \ref{fig1}a except for the position of the $PCR$ point and the slope of the invariant line. In the panel \ref{fig2}c, the inhomogeneous projection $\Ommq=-\alpha/w$ and $\Omeq=1-\alpha/w$ is displayed for a choice satisfying $1+w+\alpha<0$. The point $PC4$ is a saddle point and the point $PC6$ is now the attractor of the system, in contrast with the case $1+w+\alpha<0$ that will be similar to the quintessence and cosmological constant cases. Finally, the inhomogeneous subspace with $\Ommq=1,\,\Omeq=0$ is as in the previous case similar to that in \ref{fig2}d.

\subsection{Energy density flow from CDM to DE ($\alpha<0$).}

When $\alpha<0$, the energy flows from the CDM to DE. In this case only $PC1$, $PC2$ and $PC3$ have physical meaning while the rest of the points represent values with $\Ommq<0$. In the homogeneous subsystem, the future attractor $PCA$ is no longer physical and consequently the invariant line is also non physical. The trajectories in the homogeneous subsystem evolve from the past attractor to the $\Ommq=0$ axis or to infinity. When a given shell reaches the $\Ommq=0$ axis, we can assume that the CDM content of that shell has been consumed by the coupling term. Such points occur at different values of $\xi$ (and thus different cosmic times). The evolution of the mixture is only physically meaningful up to these points.

There is no significative difference between the homogeneous space for the different possibilities of the parameter $w$. Although the trajectories follow a different curve for every choice of $w$ and $\alpha$, they all evolve form the critical point $PCR$. Panel \ref{fig1}b shows schematically the homogeneous subspace for $\alpha<0$. The behavior of $PC1$, $PC2$ and $PC3$ in the inhomogeneous subspace with $\Ommq=1, \Omeq=0$ is identical to the $\alpha>0$ case, plotted in the panel d of figure \ref{fig2}.

The lack of future attractor for the $\delta^A$ functions ($A=m,e, \HH$, phase space variables of the inhomogeneous subspace) in this case makes it possible for some of their initial profiles to make them evolve: to infinity (shell crossing), or to some $\delta^{A}<-1$. From their definition in (\ref{qmaps}) the values $\delta^{A}=-1$ imply $A=0$ if $A_q \neq 0$ (we assume $A=\rhoqm,\,\rhoqe,\, \HH$ to be positive). Consequently, it is not possible to keep an evolution with a $\delta^A$ for more negative values than the limit $-1$, as this would imply negative local densities. We can argue that the evolution equations yield unphysical conditions if (somehow) $\delta^m,\,\delta^e<-1$ holds. In the next section, we explore this problem for a specific initial profile.

\section{Numerical example of idealised structure formation scenarios.}\label{numerical}

In this section we examine various numerical examples of evolving radial profiles that could lead to potentially interesting structure formation scenarios. Our intention is not to model any type of realistic configuration, but to illustrate the dynamical evolution of the mixtures through simple idealised examples. We choose an appropriate form for $R_0(r)$ defined for an interval $r \in [0,r_{\hbox{\tiny{max}}}]$ and examine the evolution equations for fixed values of $r$ specified by a partition of $n$ elements in this interval.

To look at the numerical evolution of the initial profiles we define the dimensionless time parameter (different from $\xi$) given by $\bar t=H_s t$ where $H_s$ is an arbitrary constant with time inverse dimensions (in cosmological applications it is customary to choose $H_s=H_0$). This rescaling of time introduces a rescaling of the remaining variables:  $\bar \HH_q=\HH_q/H_s,\,\,\kappa\bar\rhom/3= \kappa\rhom/(3H_s^2),\,\,\kappa\bar\rhoe/3= \kappa\rhoe/(3H_s^2)$. For  simplicity we will drop the bars on the normalised variables and will set the arbitrary scale as $H_s=1$, which fixes the energy density normalisation scale as $\kappa/(3H_s^2)=1$ (see \cite{izsuss17}).

In order for the initial profiles to define a structure formation scenario we need some of the ``inner'' shells (values of $r$ around the symmetry centre $r=0$) that initially expand, but at some $t$ bounce and collapse ($\HH_q$ changes sign from positive to negative), whereas ``outer'' shells continue expanding ($\HH_q>0$ holds for all $t$). The bounce is defined by $\HH_q=0$, hence we can define for each $r$ a value $t=t_{\hbox{\tiny{max}}}(r)$ such that $\HH_q(t_{\hbox{\tiny{max}}}(r),r)=0$. Notice that the dynamical systems study we have undertaken does not examine phase space trajectories of shells that have bounced and then collapse ($\HH_q\to 0$ evolving towards $\HH_q< 0$), as both coordinates $[\Ommq,\,\Omeq]$ of the homogeneous projection diverge as $\HH_q\to 0$. The numerical study given in this section will compensate for this deficiency.

For the mixed expanding/collapsing type of evolution described above we need the following homogeneous subspace trajectories: (i) the outer ever expanding shells must evolve from the past to the future attractor (or to the $\Omeq$ axis when $\alpha<0$); (ii) inner shells must evolve from the past attractor to infinity $\Omeq,\,\Ommq \rightarrow \infty$ as $t\to t_{\hbox{\tiny{max}}}$. For a bounce/collapse regime (and pending on specific initial conditions), the variables of the inhomogeneous subspace could also diverge or not evolve to the future attractor.

As mentioned before, if $\delta^A=-1$ on a given shell and $A_q\ne 0$, then $A=0$, which for positive definite quantities (densities) implies that the LTB dynamical yield an unphysical evolution for decreasing $\delta^A<-1$. This problem tends to occur specially in the cases with $\alpha<0$, when no physical attractor is present, but it may also occur for some configurations in the $\alpha>0$ cases, where the inhomogeneous attractor is physical but the initial profiles were set with the initial $\delta^A$ functions out of its attraction basin.

We have chosen the following initial profiles to be used to probe the models for three different sets of free parameters $\alpha,\,w$
\ba
\rhomi&=&{ m_{10}}+{\frac {{ m_{11}}-{ m_{10}}}{1+\tan^3(r)}},\qquad { m_{10}
}= 0.00,\quad { m_{11}}= 9.10;\nonumber\\
\rhoei &=& 0.55;\label{p15}\\
\KK_0&=&k_{10}+\frac{k_{11}-k_{10}}{1+\tan^2(r)},\qquad k_{10}=-4.10, \quad k_{11}=7.50;\nonumber
\label{struc}
\ea
together with the coordinate choice $R_0(r)=\tan(r)$. Hence, we consider a partition of $n=20$ elements for $r$ going from $0$ to $\pi/2$.

\subsection{Positive $\alpha$ and $1+w+\alpha>0$}
\begin{figure}[tbp]
\includegraphics*[scale=0.30]{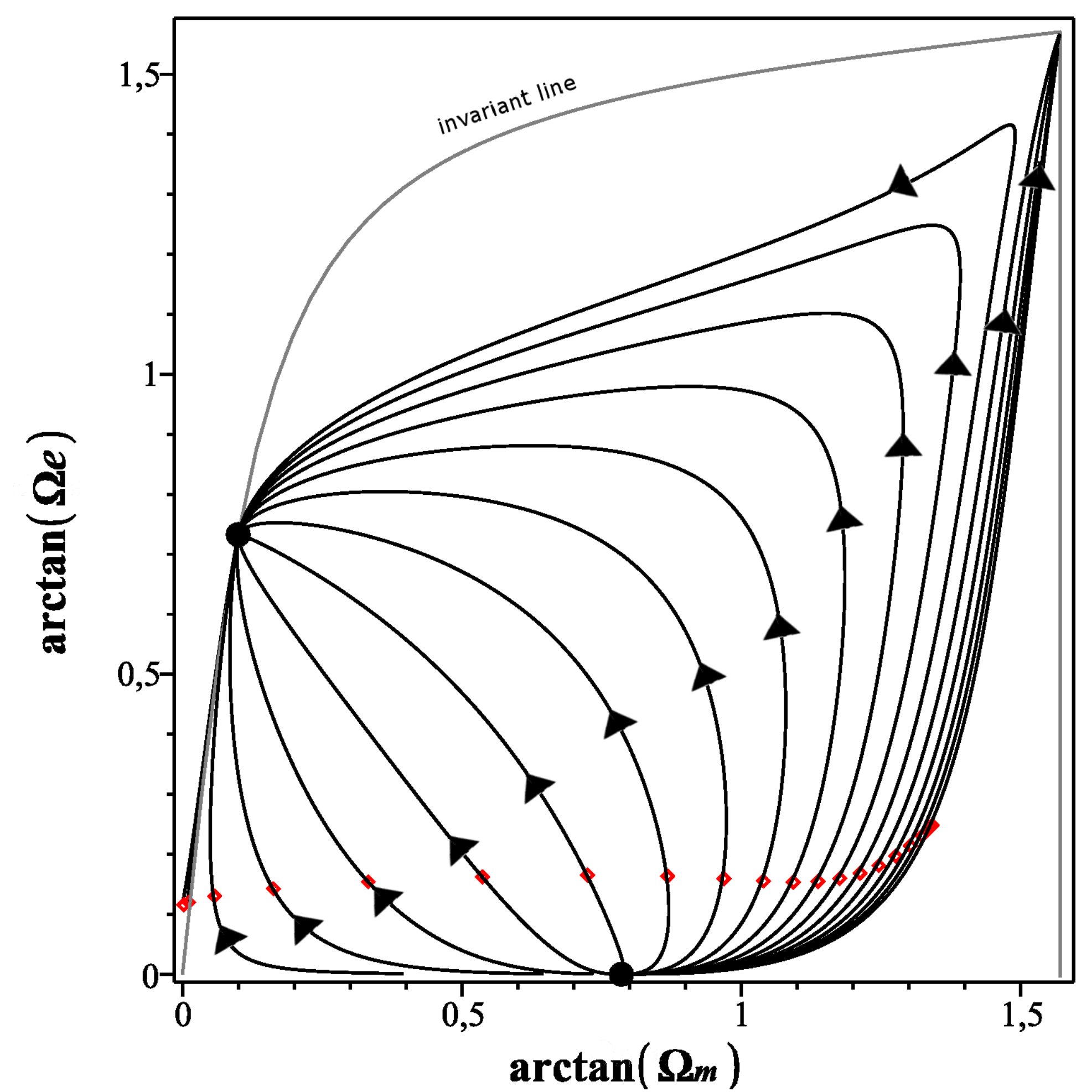}
\includegraphics*[scale=0.4]{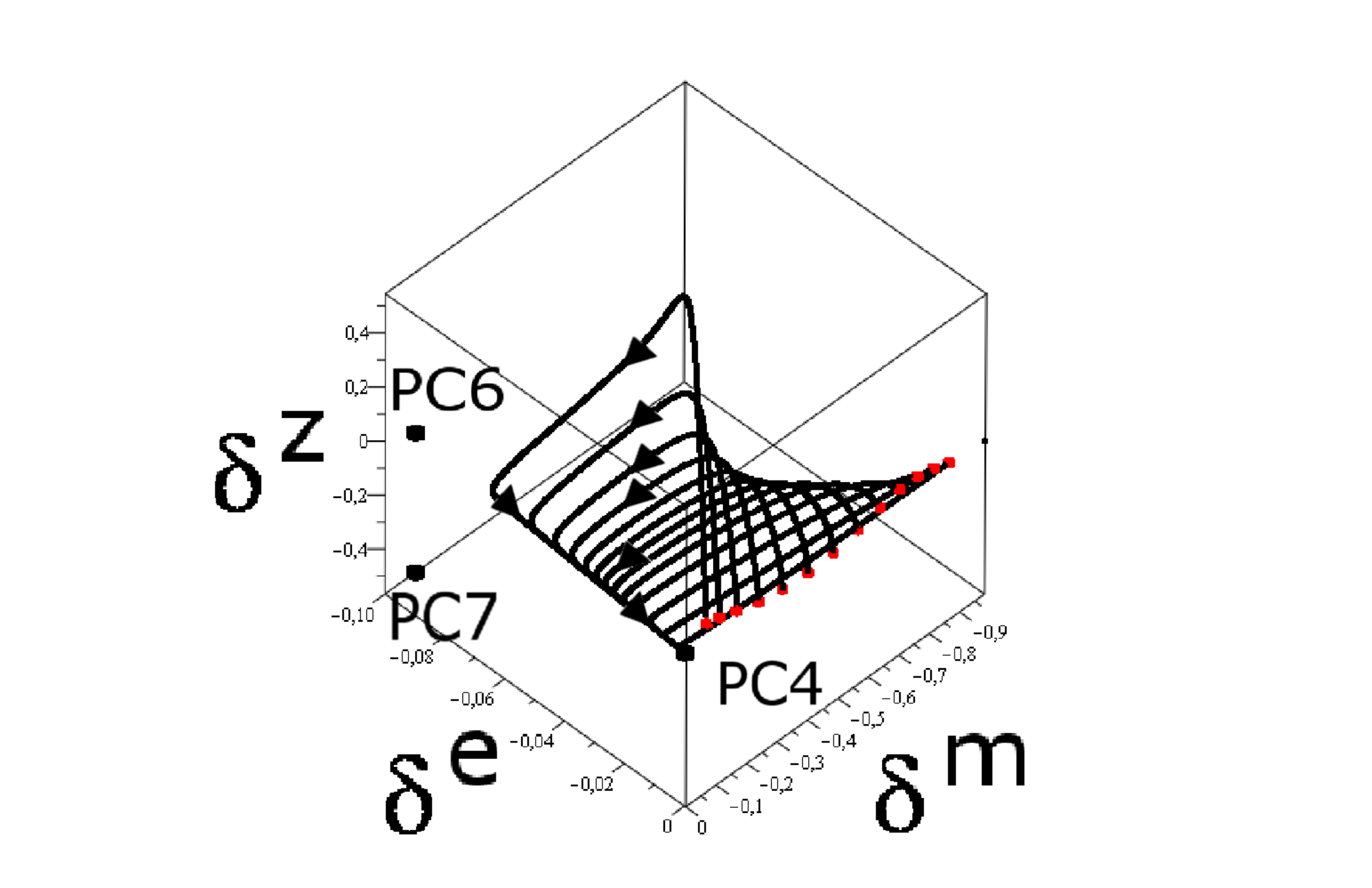}
\caption{Panel (a): Homogeneous projection of the trajectories of the system (\ref{sistdinint2a}-\ref{sistdinint2e}) for the different shells of the configuration with initial conditions given by (\ref{struc}) and $w=-1.0$, $\alpha=0.1$. The grey line represents the invariant line for this choice of parameters. The points represents the initial values of $\Ommq$ and $\Omeq$ for each shell $r=r_i$. Panel(b): Inhomogeneous trajectory of the system for the different shells with initial conditions represented as red dots. Refer to the text for a detailed discussion of the panels.} \label{fig3}\end{figure}

Considering the numerical values $\alpha=0.1$ and $w=-1$ together with the configuration (\ref{struc}), the shells $r_{1-6}$ of the partition collapse while the rest evolve to the future attractor $PC4$, as the initial values for all the shells are in the attraction basin of $PC4$. Figure \ref{fig3}a shows the homogeneous projection of the configuration with the invariant line of the system in grey, and the initial conditions for every shell as red points. Figure \ref{fig3}b shows the inhomogeneous projection of the trajectories, and the initial conditions as red points.

\begin{figure}[tbp]
\includegraphics*[scale=0.3]{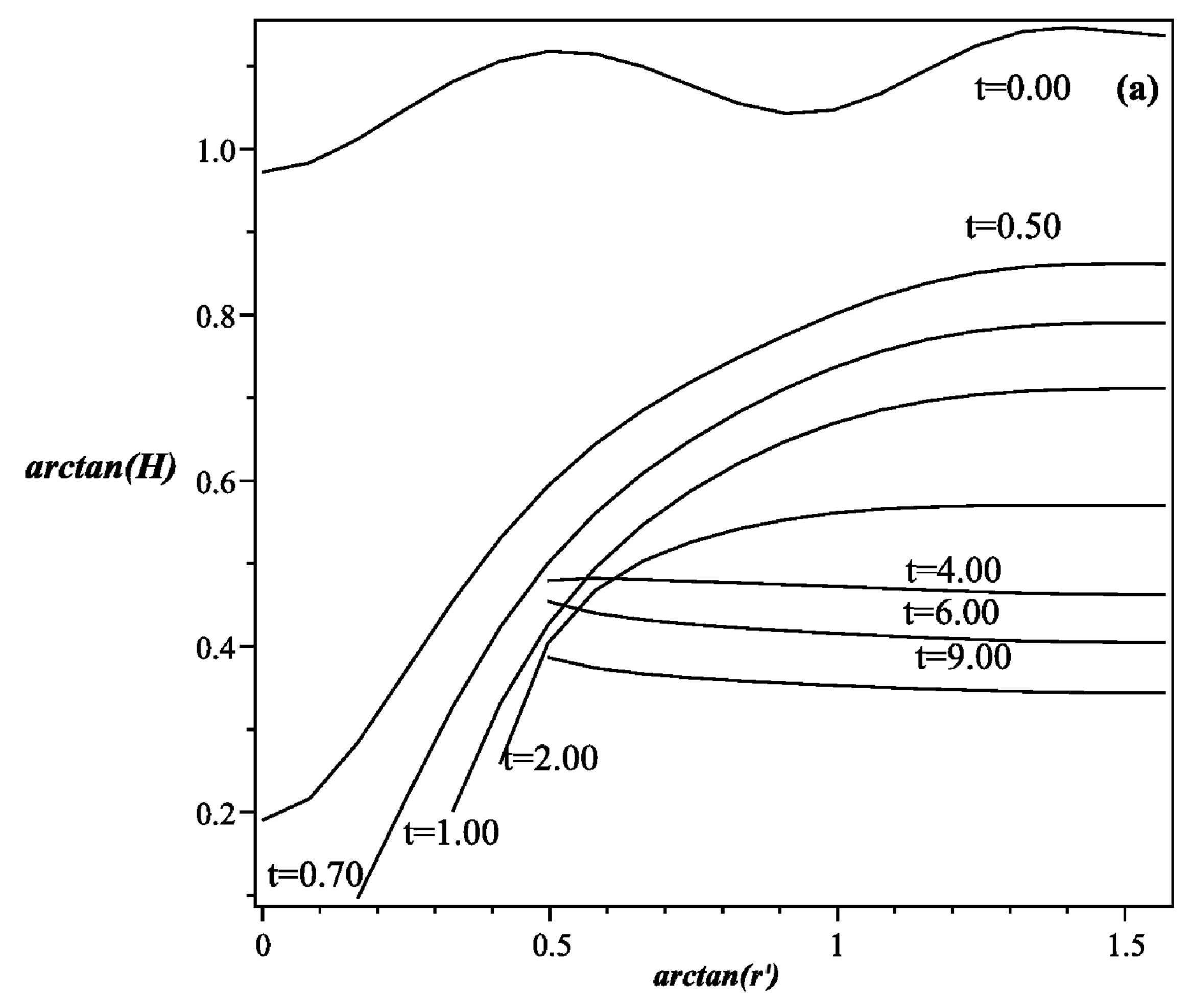}
\includegraphics*[scale=0.3]{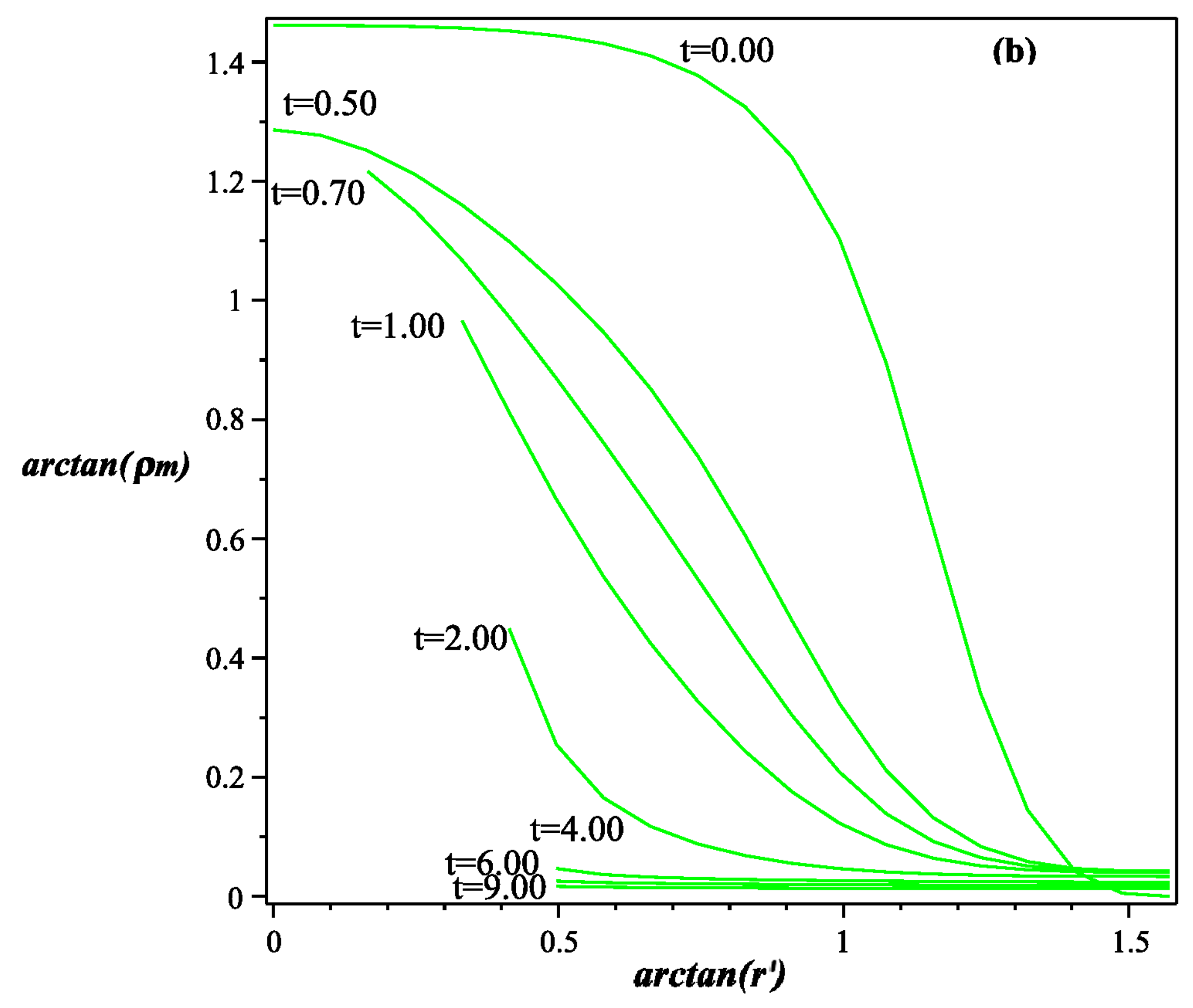}
\includegraphics*[scale=0.3]{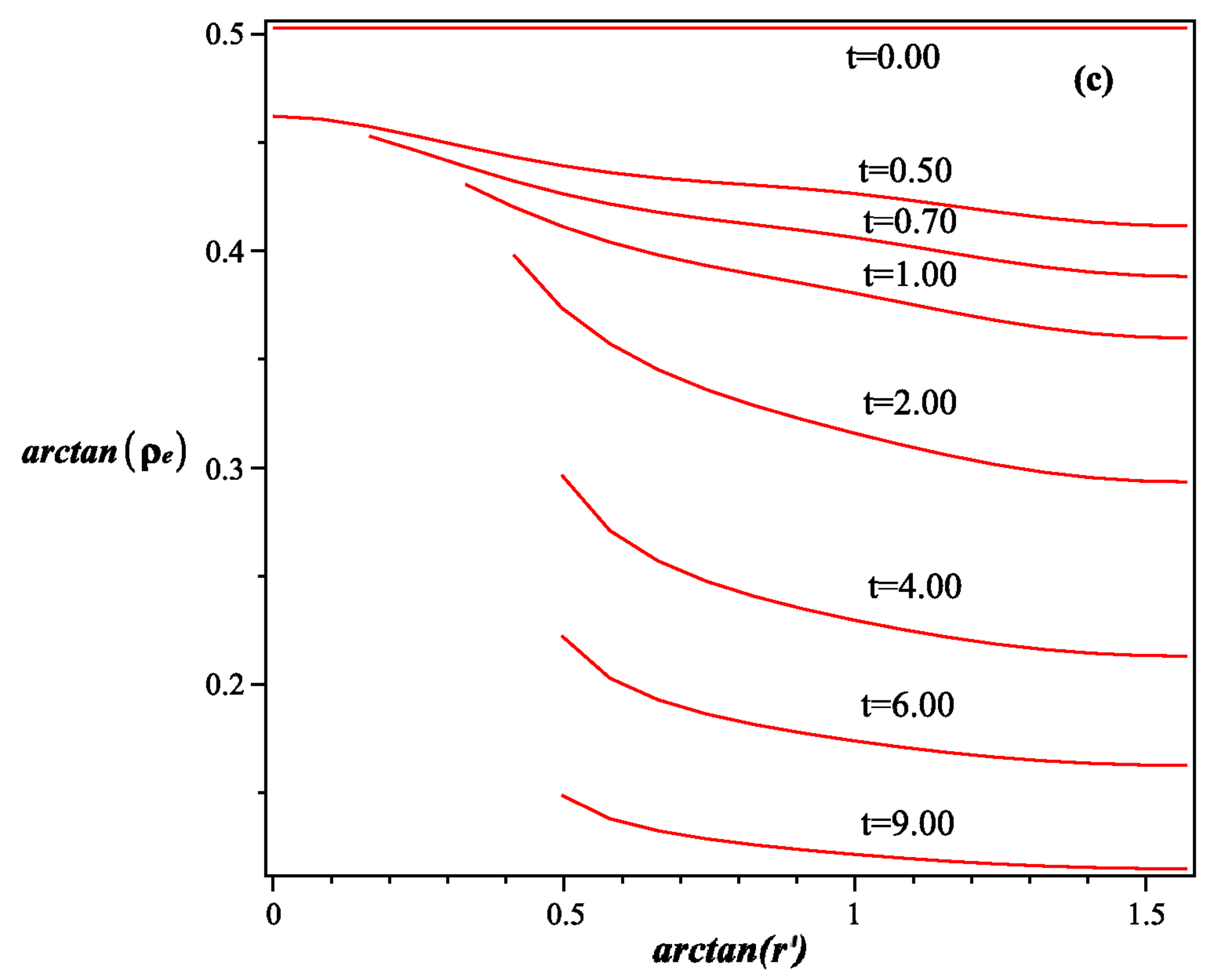}
\includegraphics*[scale=0.3]{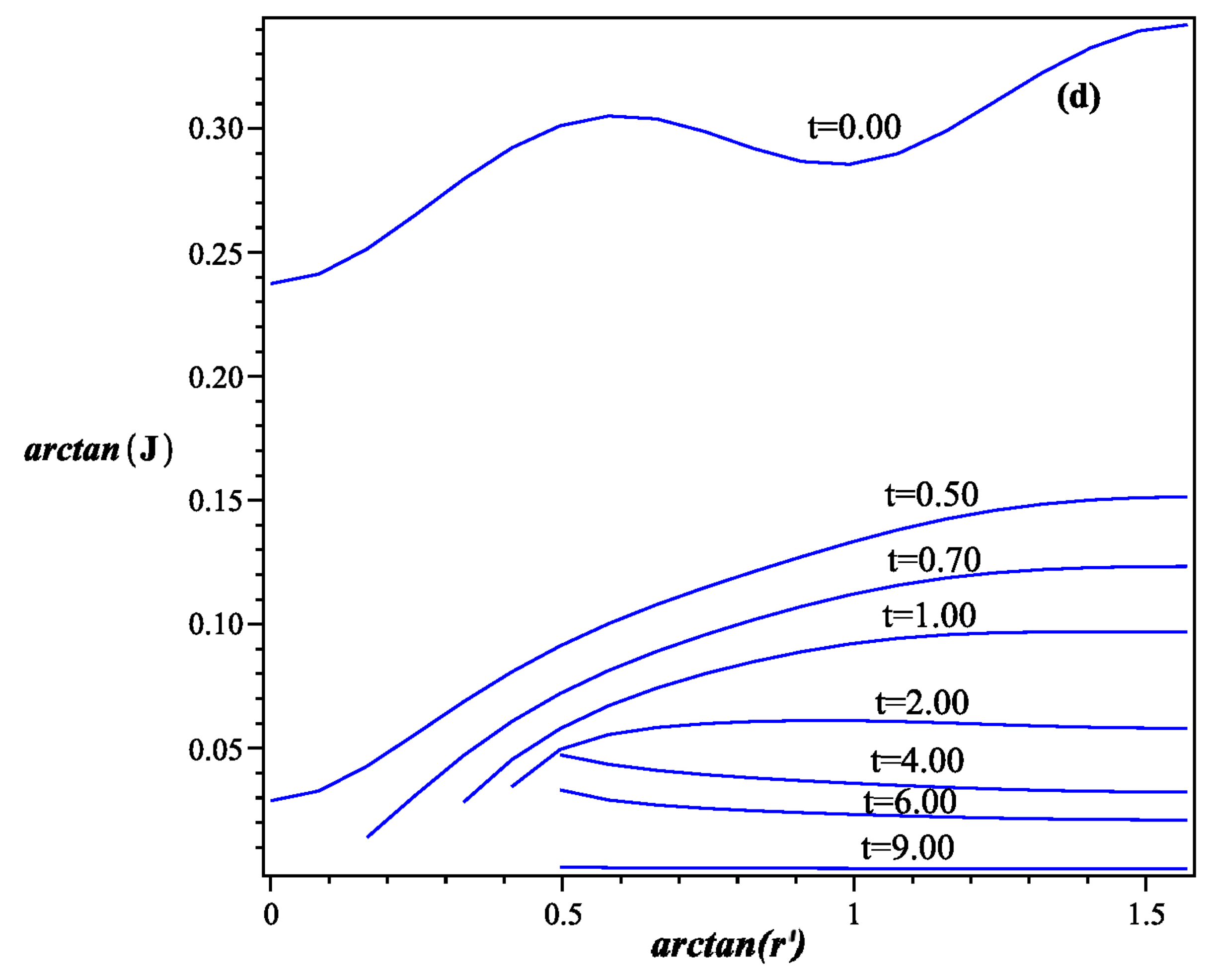}
\caption{Local profiles of different scalar functions for the configuration with initial conditions given by (\ref{struc}) and $w=-1.0$, $\alpha=0.1$ at different instants of time. Panel (a): local scalar $\HH$. Panel (b): local scalar $\rhom$. Panel (c): local scalar $\rhoe$. Panel (d): scalar $J$. Refer to the text for a detailed discussion of the panels.} \label{fig4}\end{figure}

Figure \ref{fig4} displays the radial profiles of the local scalars $A=\HH,\,\rhom,\,\rhoe$ and $J$ at different instants of time in the plane $\arctan(A)$ vs. $\arctan(R_0(r))$. For $t=0.50$ no shell has collapsed yet, for $t=0.70$ shells $1,2$ have collapsed, for $t=1.00$ the shells $3,4$ have collapsed, for $t=2.00$ shell $5$ has collapsed and the last inner shell is about to collapse. Finally, for $t=4.00$ all inner shells have already collapsed and outer shells evolve into a homogeneous profile. For expanding trajectories the q--scalars and the fluctuations $\delta^A$ for the outer shells tend to their attractor values, while the profiles of local scalars $A$ tend to a constant profile, as $\delta^{A}\to 0$ ($A=m,e,\HH$) for the attractor $PC4$. It takes a long time for the functions to evolve into their attractor values. The evolution of the these profiles to a constant profile can be appreciated in panel \ref{fig4}a for the local scalar $\HH$ at $t=4.00,6.00,9.00$ and in the panel \ref{fig4}b for the local scalar $\rhom$ at the instants $t=4.00,6.00,9.00$. The local scalar $\rhoe$, which was initially constant, needs an even longer time to evolve into a constant profile, but the line representing it at $t=9.00$ is clearly more homogeneous at the outer shells than in previous instants.

\subsection{Positive $\alpha$ and $1+w+\alpha<0$}
We choose  $\alpha=0.1$ and $w=-1.15$. Only the shells $r_{1-3}$ collapse, while the rest evolve towards the attractor $PC6$. As for the other choice of parameters, the initial $\delta^A$ functions are in the attraction basin of $PC6$. In this case, the inhomogeneous projection of the attractor is not zero as $\Dm=\De=0.05, \Dh=0.05/2$. Consequently, from (\ref{qmaps}), the profiles of $\rhom$ and $\rhoe$ do not evolve to a constant profile as in the previous case, but to a profile whose $r$--dependence is given by $\rhom=\rhoe=R^{0.15}(1.05)$, while the local $\HH$ is given by $\HH=R^{0.07}(1+0.05/2)$.

\begin{figure}[tbp]
\includegraphics*[scale=0.30]{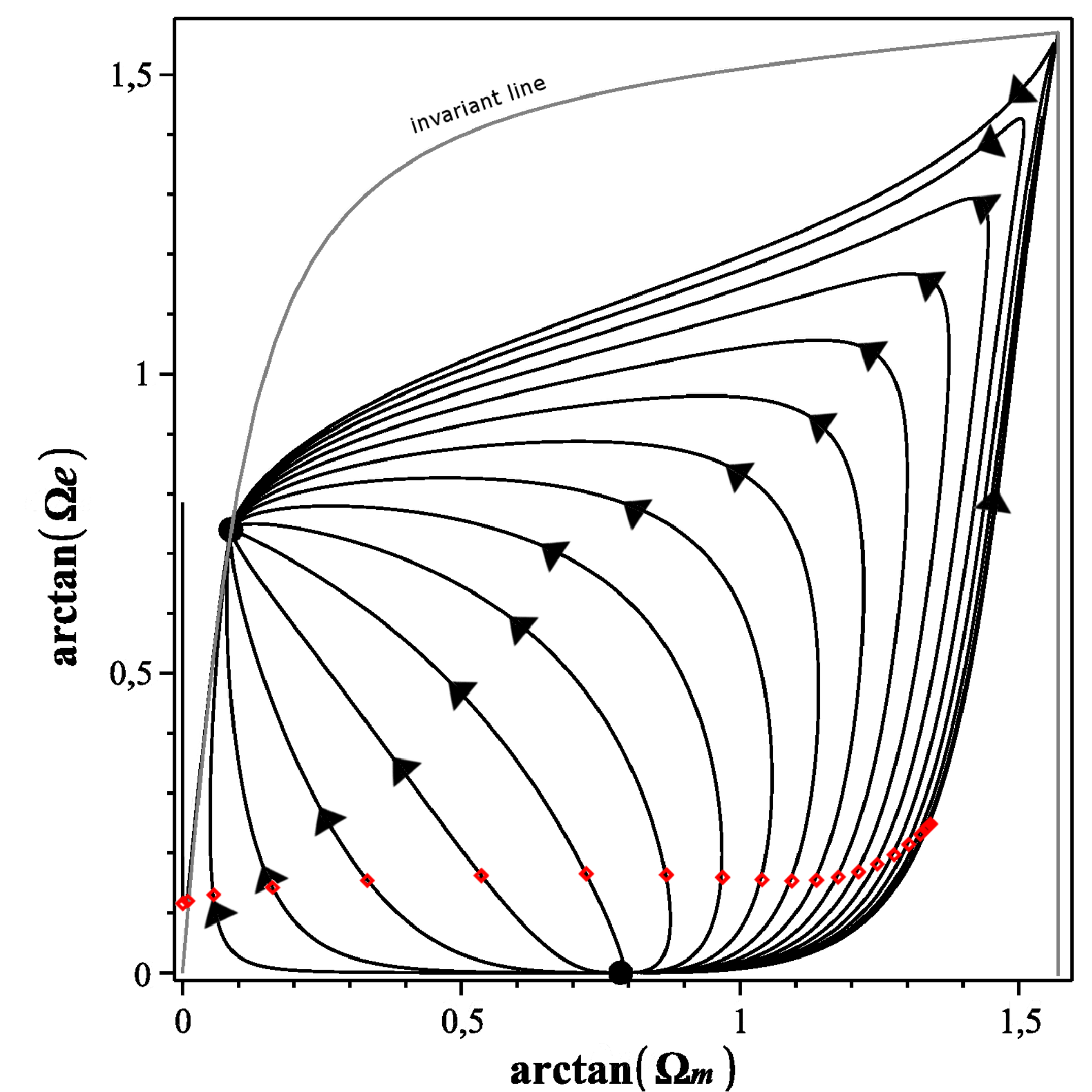}
\includegraphics*[scale=0.40]{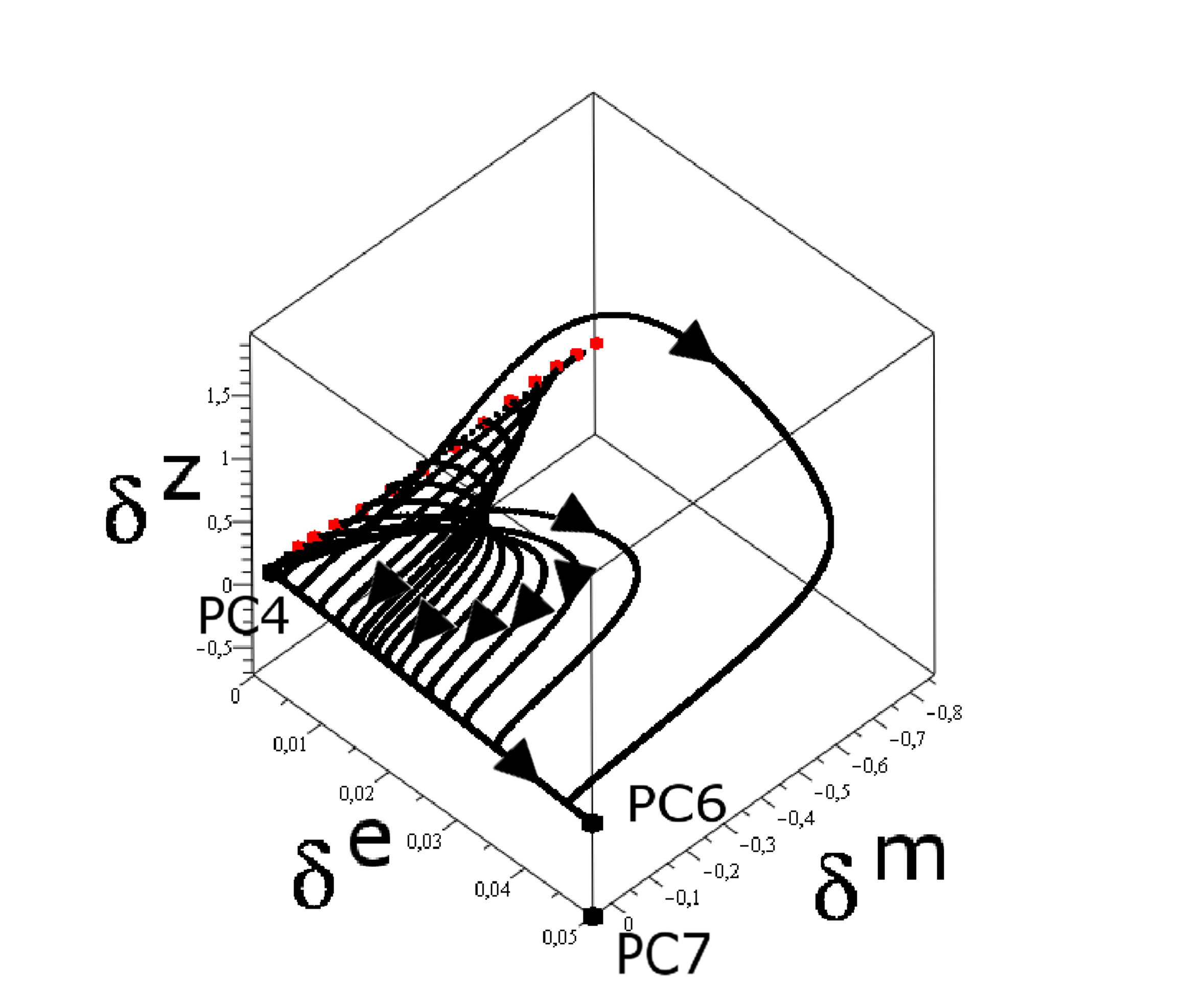}
\caption{Panel (a): Homogeneous projection of the trajectories of the system (\ref{sistdinint2a}-\ref{sistdinint2e}) for the different shells of the configuration with initial conditions given by (\ref{struc}) and $w=-1.0$, $\alpha=0.1$. The grey line represents the invariant line for this choice of parameters. The points represent the initial values of $\Ommq$ and $\Omeq$ for each shell $r=r_i$ of the partition.  Panel(b): : Trajectory of the system for the different shells with initial conditions represented as red dots. Refer to the text for a detailed discussion of the panels.} \label{fig5}\end{figure}

\begin{figure}[tbp]
\includegraphics*[scale=0.3]{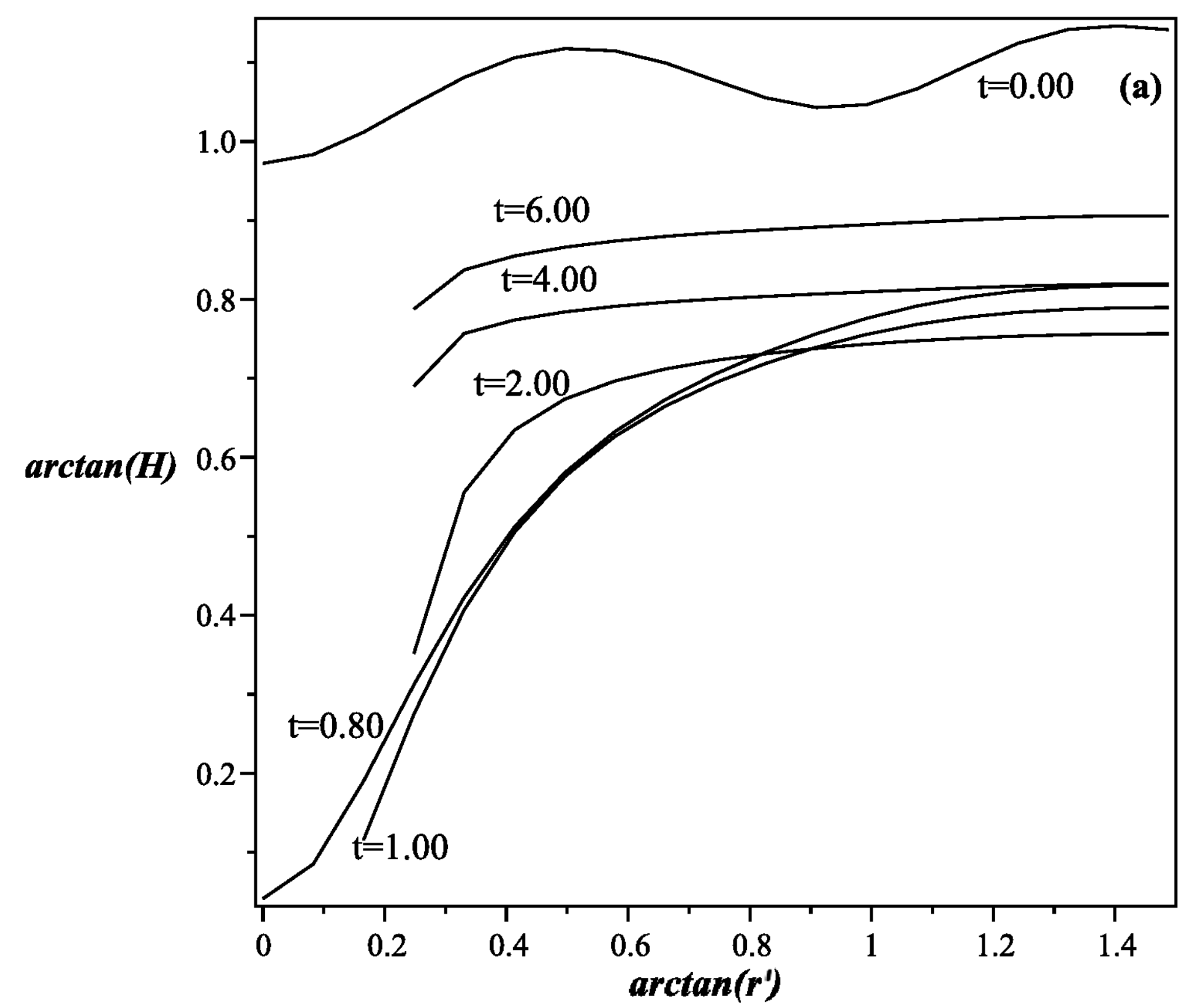}
\includegraphics*[scale=0.3]{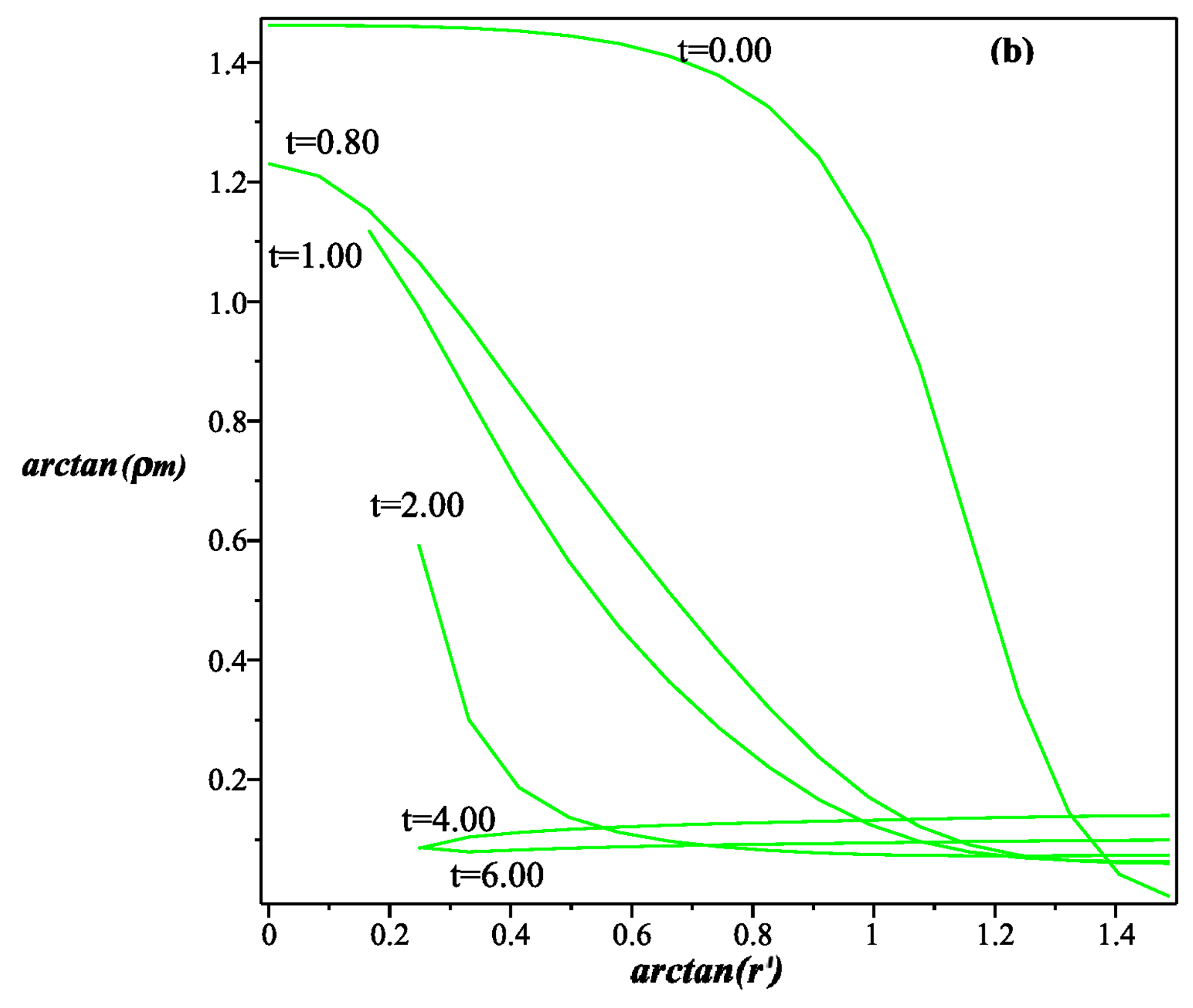}
\includegraphics*[scale=0.3]{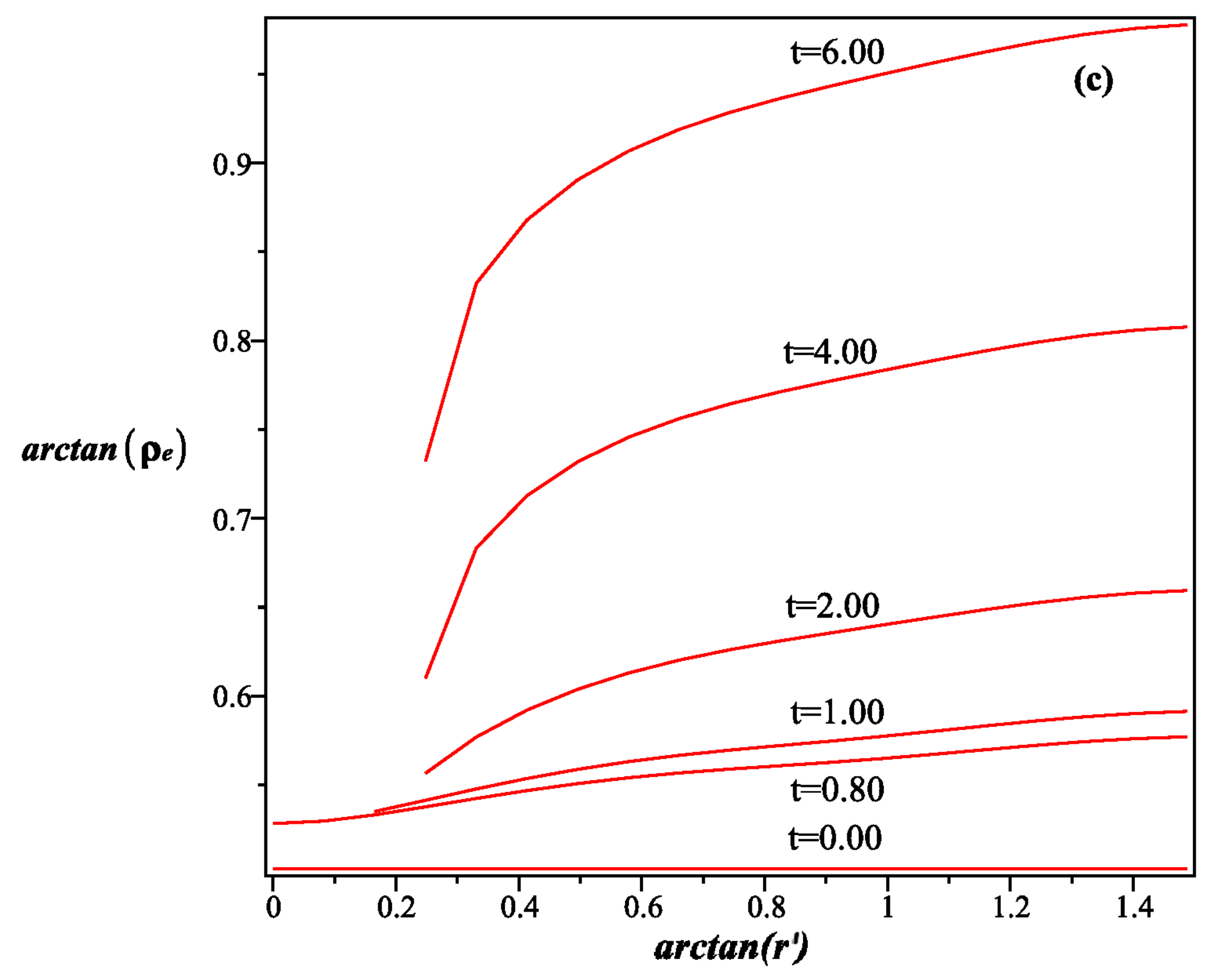}
\includegraphics*[scale=0.3]{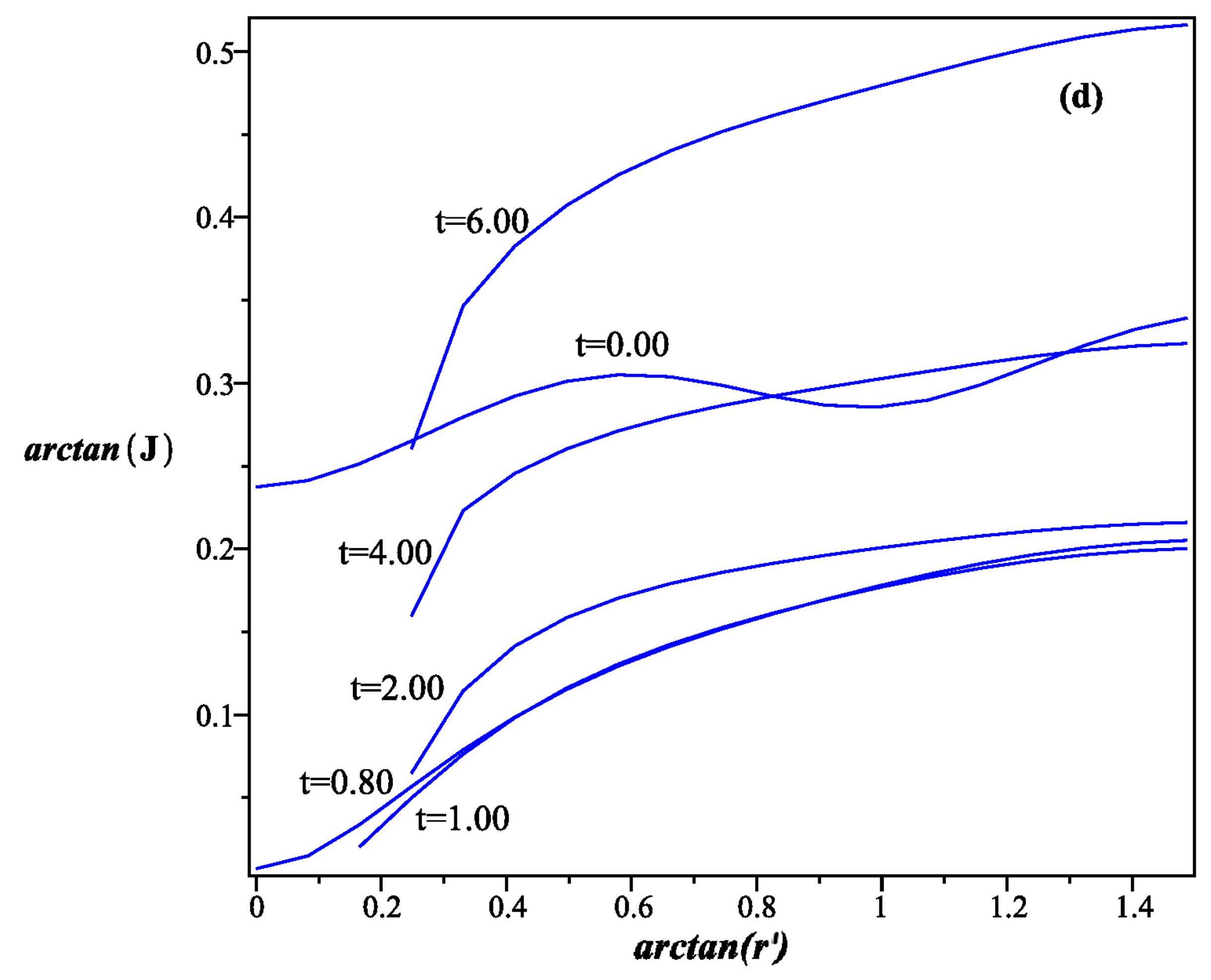}
\caption{Local profiles of different scalar functions for the configuration with initial conditions given by (\ref{struc}) and $w=-1.15$, $\alpha=0.1$ at different instants of time. Panel (a): scalar $\HH$. Panel (b): scalar $\rhom$. Panel (c): scalar $\rhoe$. Panel (d): scalar $J$. Refer to the text for a detailed discussion of the panels.} \label{fig6}\end{figure}

Figure \ref{fig5}a shows the homogeneous projection of the configuration with the invariant line of the system in grey, and the initial conditions for every shell as red points. Figure \ref{fig5}b shows the inhomogeneous projection of the trajectories evolving to the future attractor, and the initial conditions as red points.
Figure \ref{fig6} displays profiles of local scalars at different instants of time. At the instant $t=0.80$ no inner shell has collapsed yet, at the instant $t=1.00$ shells $i=1,2$ have already collapsed and at the instant $t=2.00$ all the inner shells have collapsed. Since the q--scalars and fluctuations $\delta^A$ for the outer shells tend to their attractor values while trajectories expand, the  profiles of local scalars tend to the terminal profile shown in panel \ref{fig6}a for $\HH$ at $t=4.00,6.00$ and in the panel \ref{fig6}c for $\rhoe$ at instants $t=4.00,6.00$.

\subsubsection{Negative $\alpha$}
For any choice of $\alpha<0$, the future attractor is no longer physical. To illustrate this case, we chose $\alpha=-0.1$ and $w=-1$. The shells $r_{1-4}$ collapse while outer shells, $r_{5-20}$, keep their expanding evolution up to a point where $\Dm =-1$ and the LTB evolution is no longer physical. In particular for this profile and this choice of parameters the function $\Dm$ tends to $-1$ very rapidly for the outer shells.

\begin{figure}[tbp]
\includegraphics*[scale=0.30]{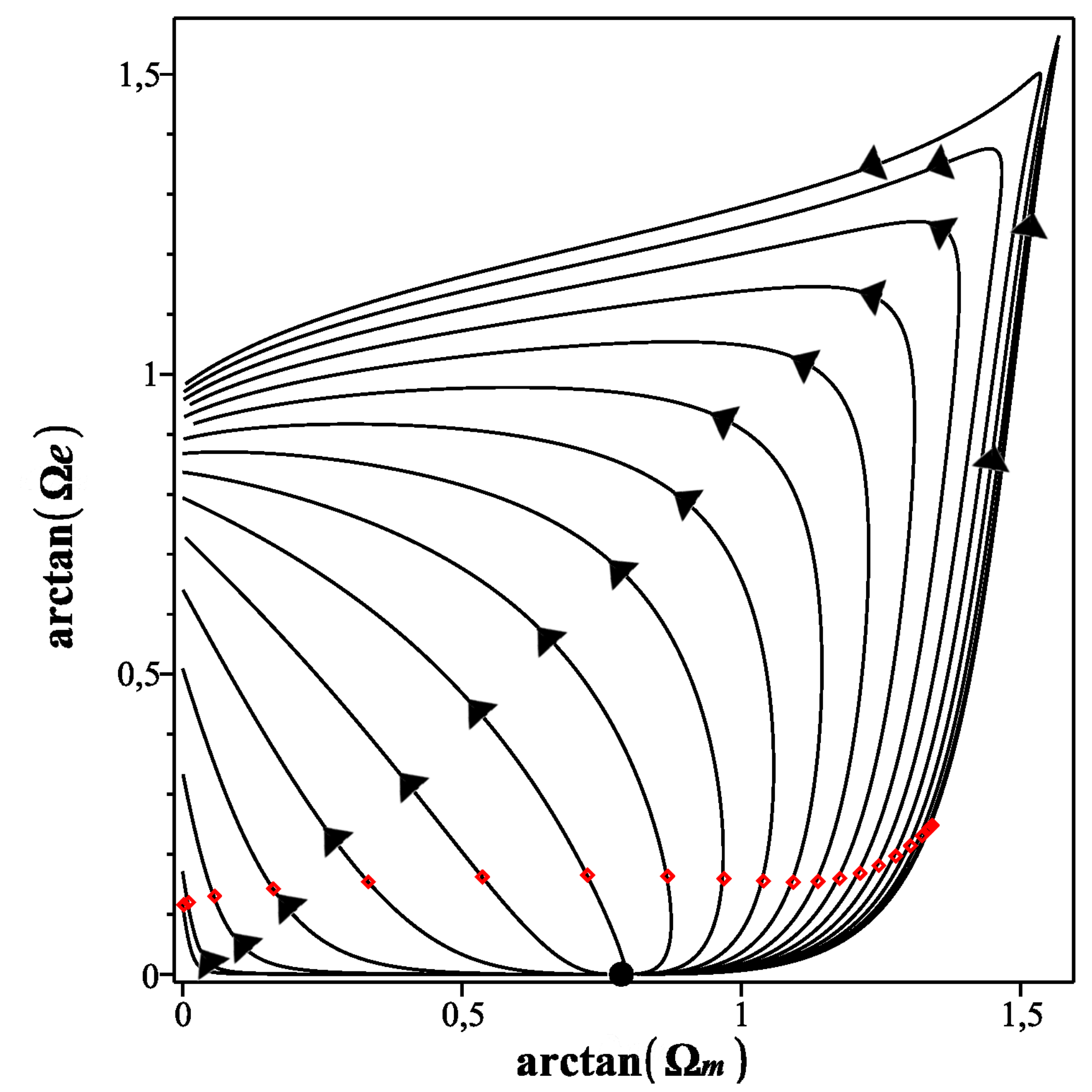}
\includegraphics*[scale=0.30]{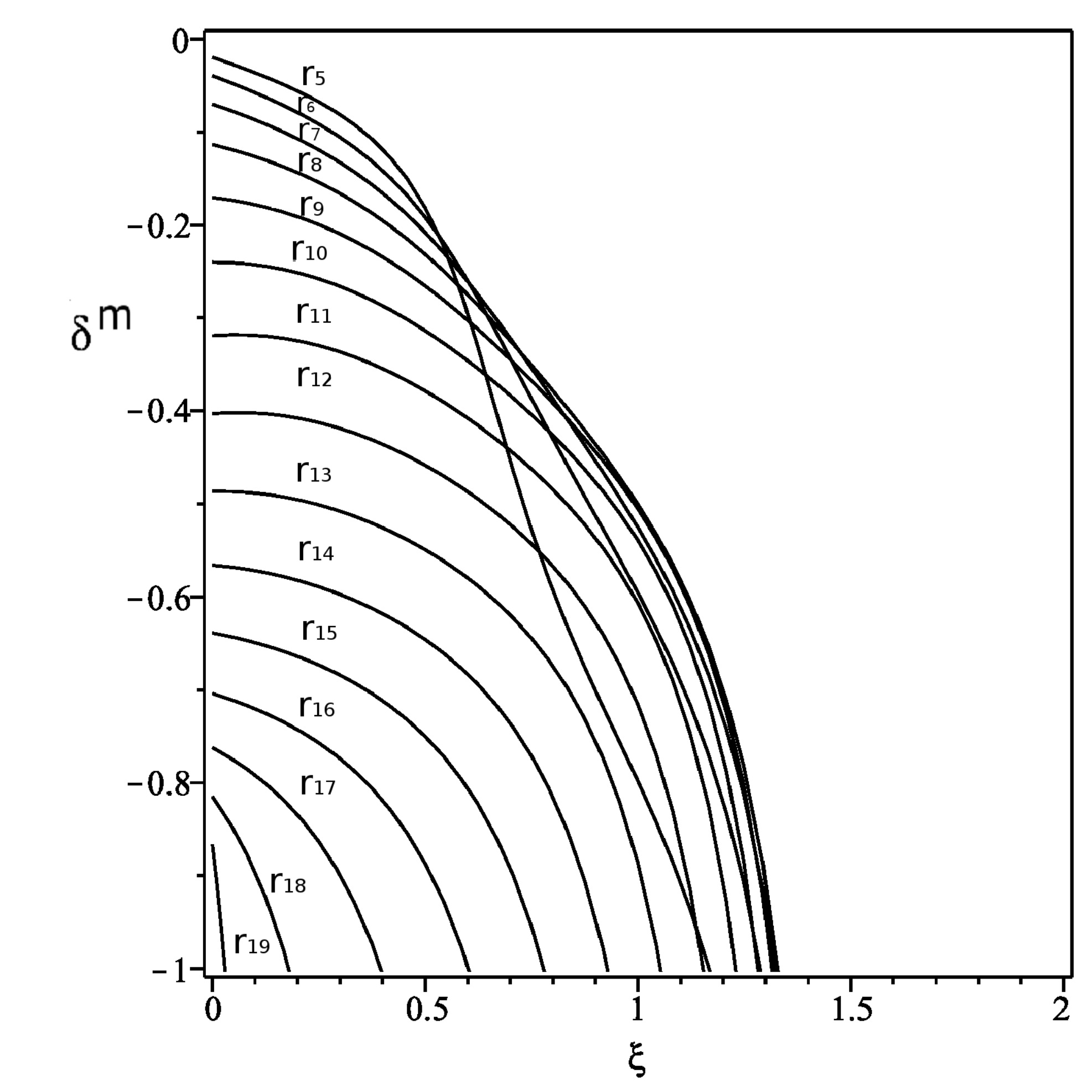}
\caption{Panel (a): Homogeneous projection of the trajectories of the system (\ref{sistdinint2a}-\ref{sistdinint2e}) for the different shells of the configuration with initial conditions given by (\ref{struc}) and $w=-1.0$, $\alpha=-0.1$. The grey line represents the invariant line for this choice of parameters. The points represent the initial values of $\Ommq$ and $\Omeq$ for each shell $r=r_i$. Panel(b): $\Dm$ vs $\xi$ for the outer shells. Refer to the text for a detailed discussion of the panels.} \label{fig7}\end{figure}

Figure \ref{fig7}a shows the homogeneous projection of the configuration with the invariant line of the system in grey, and initial conditions for every shell as red points. All the outer shells evolve to the $\Ommq=0$ axis. Figure \ref{fig7}b displays the plot $\Dm$ vs. $\xi$ for the outer ever expanding shells $i=5-19$. The first shell to reach the value $\Dm=-1$ is $i=20$, which is not represented in the figure as $\Dm\to -1$ occurs immediately for this shell. The value $\xi_{i}$ for which $\Dm=-1$ occurs ({\it i.e.} $\Dm(r=r_i,\xi_i)=-1$) puts an upper limit to the range of $\xi$ for which we can use eqs. (\ref{sistdinint2a}-\ref{sistdinint2e}) to obtain all the local scalars.

\begin{figure}[tbp]
\includegraphics*[scale=0.3]{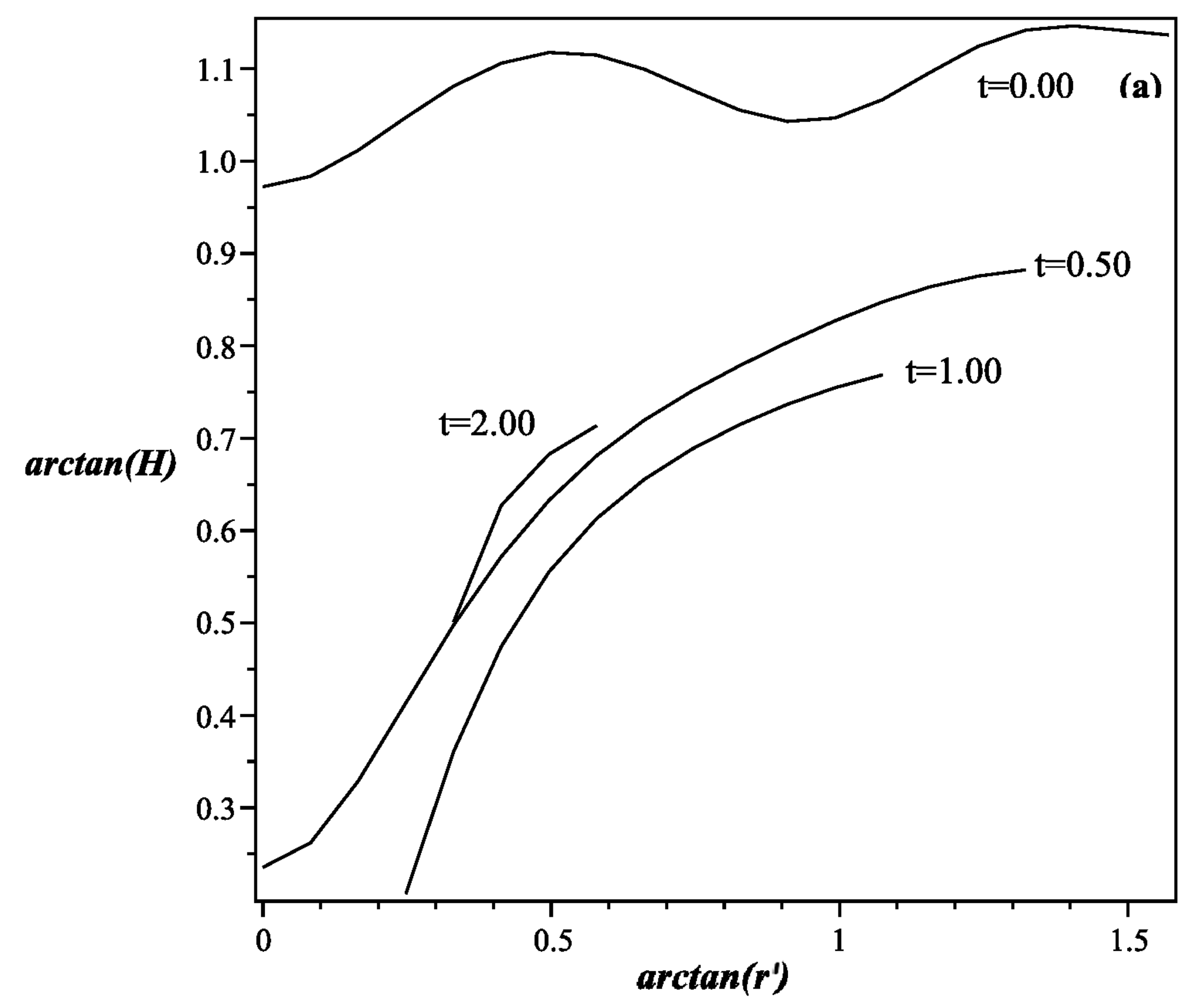}
\includegraphics*[scale=0.3]{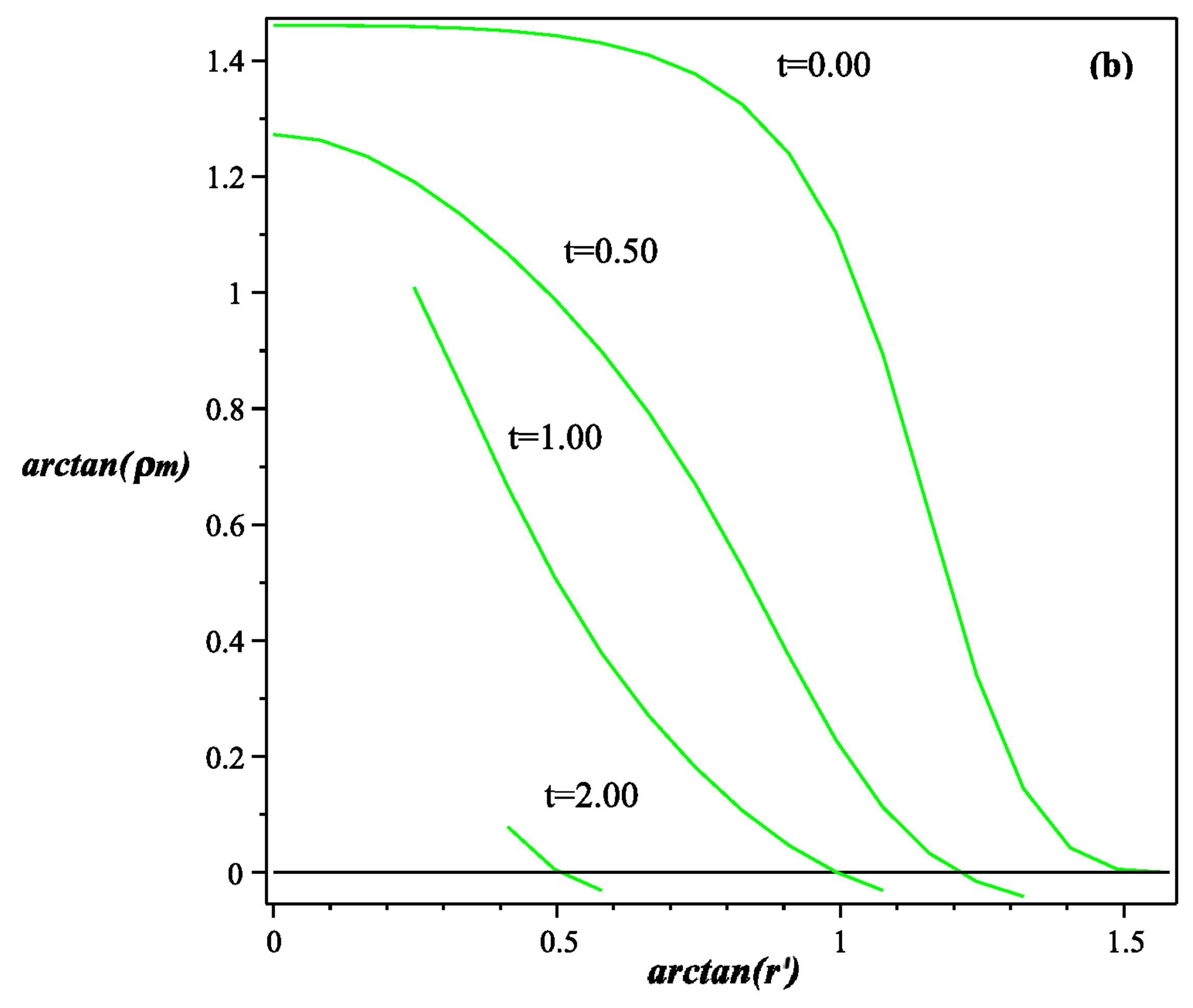}
\includegraphics*[scale=0.3]{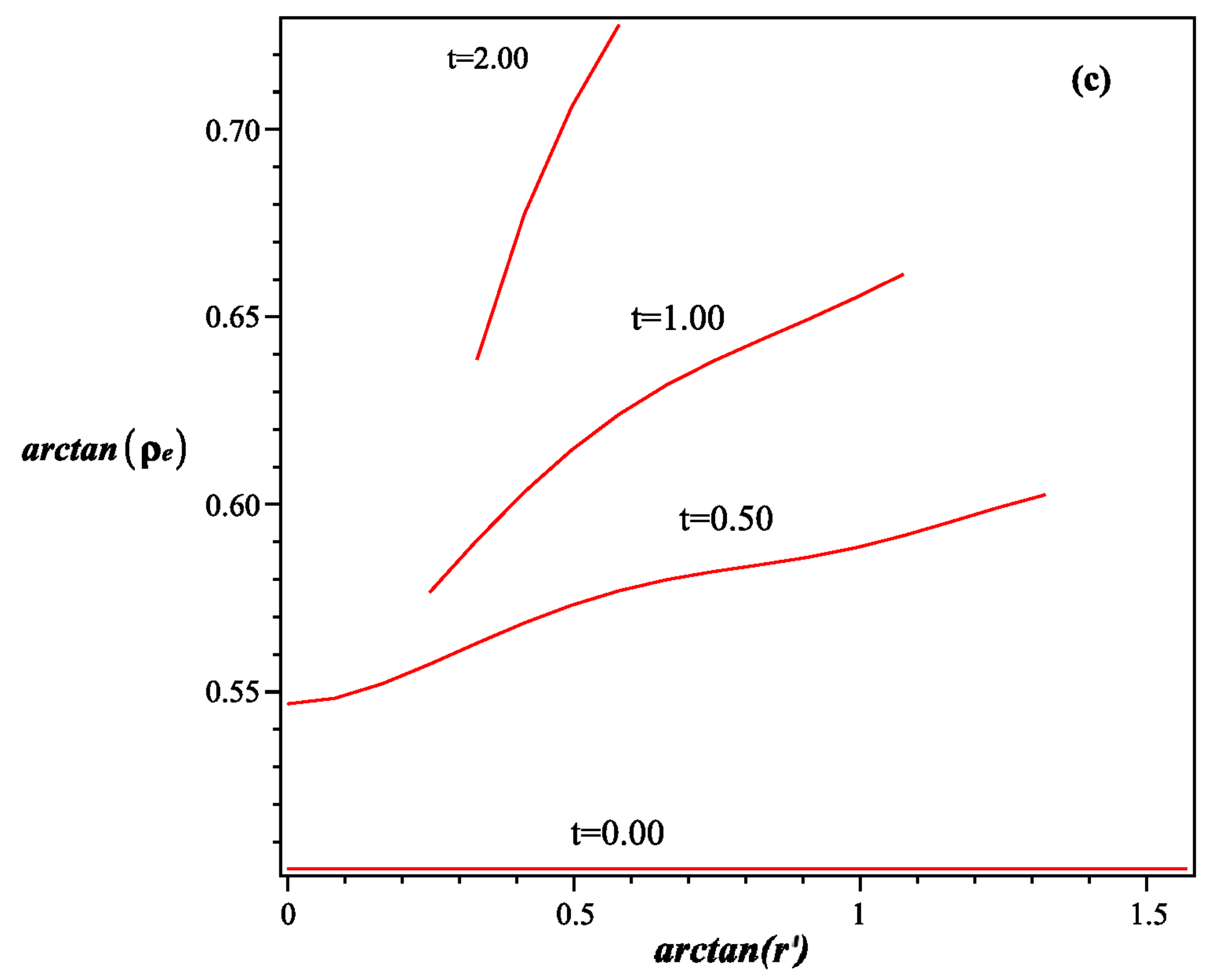}
\includegraphics*[scale=0.3]{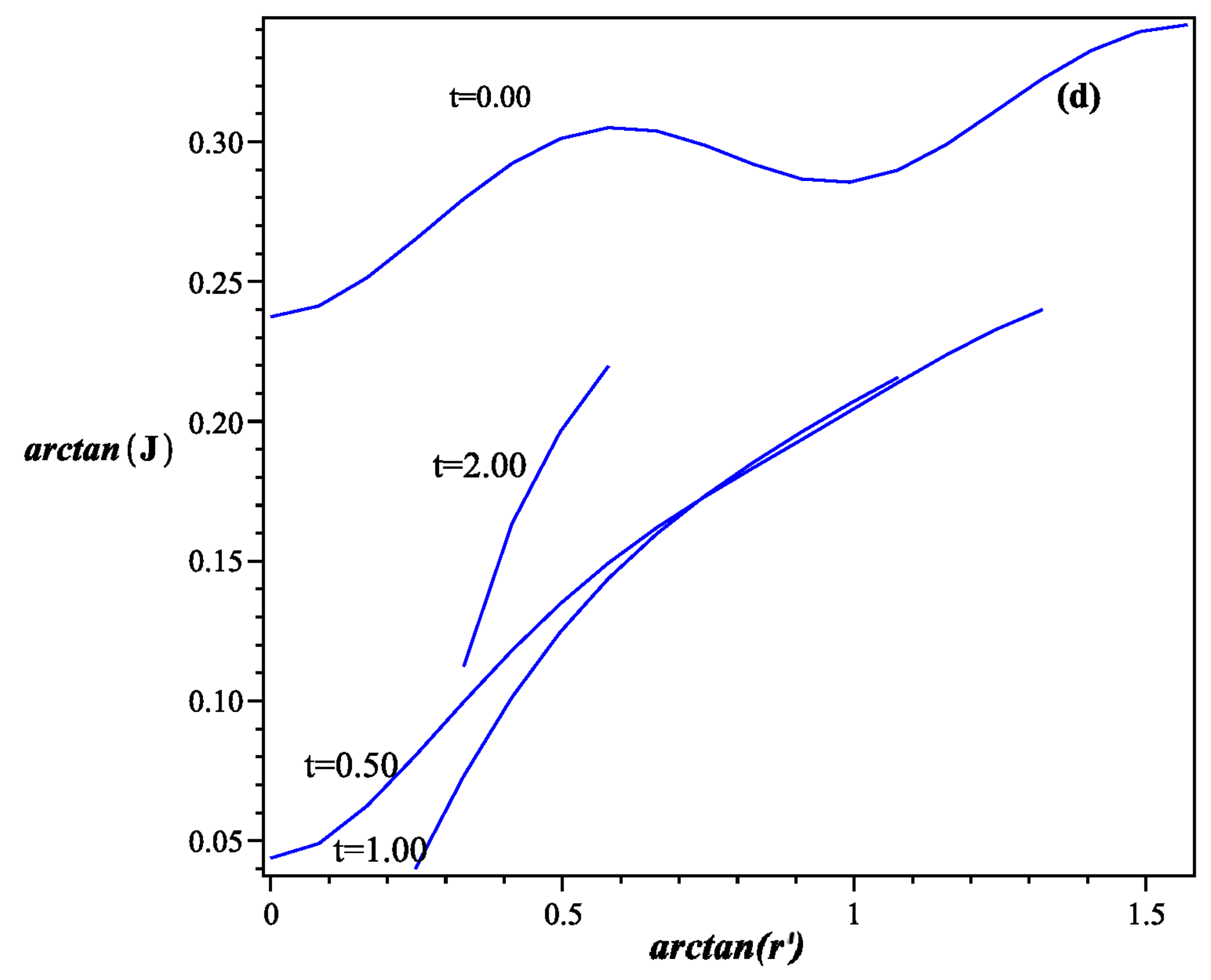}
\caption{Profiles of different local scalars for the configuration with initial conditions given by (\ref{struc}) and $w=-1.0$, $\alpha=-0.1$ at different instants of time. Panel (a): scalar $\HH$. Panel (b): scalar $\rhom$. Panel (c): scalar $\rhoe$. Panel (d): scalar $J$. Refer to the text for a detailed discussion of the panels.} \label{fig8}\end{figure}

Figure \ref{fig8} shows local profiles of various scalars at different instants of time. The profile of the local scalar $\rhom$ tends to zero for every shell at the instant mentioned earlier. This behaviour is shown in figure \ref{fig8}b for different instants of time. At the instant $t=0.50$ the shells $17-20$ display an unphysical evolution. At the instant $t=1.00$ the shells $14-17$ join the shells mentioned before and are then followed by shells $8-13$ before the instant $t=2.00$. On the other hand, the inner shells have not collapsed at the instant $t=0.50$. At $t=1.00$ the shells $1-3$ have already collapsed and the shell $4.00$ is about to follow. By the instant $t=2.00$, all the inner shells have collapsed. The rest of the panels represent the profiles of the other local scalars. In every case, the scalars of the outer shells are not computed from the instant at which the LTB evolution becomes unphysical.

\section{Conclusions.}\label{conclusions}

We have undertaken a full study of the phase space evolution of expanding and interactive CDM-DE mixtures, under the assumption that the interactive term (see (\ref{intj_e})) is proportional to the DE energy density. These mixtures are the source of an inhomogeneous and spherically symmetric exact solution of Einstein's equations characterised by an LTB metric. The present article, together with a recent article \cite{izsuss17}, generalise previous work \cite{izsuss10} for a CDM source with DE modelled by a cosmological constant.

As in \cite{izsuss10,izsuss17}, we examined the dynamics of these LTB solutions by means of q--scalars and their fluctuations, which transform in a natural way Einstein's equations into a dynamical system evolving in a 5--dimensional phase space. The latter was studied in terms of two interrelated subspace projections: the 2--dimensional homogeneous subspace whose variables are q--scalars ($\Ommq,\,\Omeq$) analogous to covariant scalars of an FLRW model with same type of CDM-DE mixture, and a 3--dimensional subspace involving the fluctuations of the q--scalars ($\Dm,\,\De,\,\Dh$) that control the inhomogeneity of the models (deviation from FLRW).

The critical points associated with the phase space are listed in  Table \ref{criticalpoints}: a past attractor, a future attractor and 5 saddle points. All of them (save the past attractor) depend on the two constant free parameters of the solutions: the EOS parameter $w$ and the proportionality between the interaction term and DE density, $\alpha$, whose sign determines the directionality of the interaction energy transfer (CDM to DE for $\alpha>0$ and DE to CDM for $\alpha<0$). The phase space evolution was examined for ``quintessence'' models ($-1<w<-1/3$) and ``phantom'' models ($w<-1$), keeping in either case $w$ close to the value -1 that is favoured by observations.

It is important to compare our results with those found in our recent study in \cite{izsuss17} involving a similar CDM-DE mixture, but with the interaction term proportional to the CDM density through the dimension--less constant $\alpha$. The main difference between this assumption and that of the present work (interaction proportional to DE density) is in the parameter dependence of the past attractor, which in both mixtures can be associated with the initial Big Bang singularity. For the mixture of  \cite{izsuss17} the phase space position of this attractor depended on  $\alpha,\,w$, while in the present mixture it does not, which means that it is a fixed point in the phase space.

The above mentioned difference in the phase space position of the past attractor has very important consequences, as all available cosmological observations survey our past light cone. For $\alpha<0$, and for every set of initial conditions, the past attractor in \cite{izsuss17} was located in an unphysical phase space region (DE density becomes negative), hence $\alpha>0$ (energy flows from CDM to DE) was the only physically plausible choice that can be (in principle) compatible with observational constraints for all choices of EOS parameter $w$. As a contrast, in the mixture examined here we found that regardless of the sign of $\alpha$ the past attractor is fixed, taking physically plausible values: an Einstein de Sitter state with zero DE density and positive CDM density with unity Omega factor ($\Omeq=0,\,\Ommq=1$). Hence, both directions of the interaction energy transfer are (in principle) compatible with observations.

Since the future attractors correspond to times much beyond the present cosmic time, they cannot be contrasted with observational data, and thus are more amenable to speculation. For the case examined in \cite{izsuss17} this attractor simply marked a fixed de Sitter state ($\Ommq=0,\,\Omeq=1$). However, in the present study the phase space position of the future attractor depends on the choice of $w,\,\alpha$, and for $\alpha<0$ this position is not in a physically meaningful phase space region (CDM density becomes negative). While this can be problematic (as all trajectories must terminate in this attractor), it can still be acceptable provided we only consider the evolution of the models up to phase space points where the CDM density vanishes. Such points correspond to values for which $\Dm=-1$ that mark different cosmic times for different comoving shells. On the other hand, for $\alpha>0$, the shells with initial conditions in the attraction basin evolve to a point where both CDM and DE reach a terminal  energy density. In this case a choice of parameters with $1+w+\alpha>0$ lead to a final homogeneous state, as the phase space variables of the inhomogeneous subspace (the fluctuations) tend to zero. For $1+w+\alpha<0$ the comoving shells evolve towards a nonzero density profile, since the fluctuations tend to nonzero values that depend on the radial coordinate.

The comparison between the results of the present work and those of \cite{izsuss17} provide what can be, perhaps, the most interesting conclusion that follows from the present article: if CDM and DE are assumed to interact, then mixtures in which the interaction term is proportional to the DE density (as in this paper) offer more possibilities for educated speculation, and thus could be preferable over those in which it is proportional to the CDM density (as in \cite{izsuss17}). The reason for this is, as explained before, that the past attractor (which determines the past evolution surveyed by observations) is always physical for any reasonable choice of parameters, in contrast with the coupling used in \cite{izsuss17} where the choice of $\alpha<0$ was not possible.

We also examined three structure formation scenarios, given (for example) by the profile in (\ref{struc}). These initial conditions correspond to inner expanding shells that are not in the attraction basin while the rest of the expanding shells (outer shells) that evolve to the attractor or to the $\Ommq=0$ axis. The inner shells will evolve to a point where they bounce and start a collapse. The outer shells expand forever and evolve into a profile determined by the choice of $1+w+\alpha$, as mentioned before. Some examples of those scenarios can be found in section \ref{numerical}, where we have computed the local scalar functions over physical time $t$.

It is also important to compare our results (and those of \cite{izsuss17}) with those obtained for DE models or similar CDM-DE mixtures in FLRW cosmologies  \cite{copeland,boe08,ol, ol2, Maar, Gavela,db,smlm}, since the dynamics of LTB solutions described by q--scalars and their fluctuations can be mapped to linear perturbation on an FLRW background \cite{suss15}. While the spherical symmetry of the LTB models allows for the description of a single structure, the evolution of the latter can be studied exactly throughout the full non--linear regime.

We believe that our results can provide interesting clues to test theoretical assumptions on dark sources (DE and CDM) in terms of observations in the scales of structure formation and in the non--linear regime. In fact, it is straightforward to generalise LTB models to non--spherical Szekeres models, which are endowed with more degrees of freedom and thus allow for a fully relativistic  description and modelling of multiple structures \cite{sussSzkeeres1,sussSzkeeres2,sussSzkeeres3}. We are currently undertaking further efforts to extend the present work to probe non--linear observational effects of theoretical assumptions on CDM and DE sources. In particular, we aim at considering ``spherical collapse models'', as well as less idealised non--spherical models, whose source is the type of CDM--DE mixtures we have examined here, but now attempting to fit  more  realistic observational constraints of structure formation.

\section*{Acknowledgments}

RAS acknowledges support from CONACYT project number CONACYT 239639 and PAPIIT-DGAPA RR107015.

\end{document}